\documentclass[amsmath,twocolumn,prb,aps]{revtex4-1}
\usepackage{amsthm,amsfonts,graphicx,verbatim, color, tabularx}
\usepackage{bm}
\usepackage[T1]{fontenc}
\usepackage[utf8]{inputenc}
\usepackage{physics}
\usepackage[colorlinks=true, linkcolor = blue, citecolor = blue, urlcolor = blue]{hyperref}

\newcommand{\mb}[1]{\ensuremath{\mathbf #1}}

\newcommand{\dimension}{\ensuremath{\mathcal D}}

\graphicspath{{./long_figs/}}

\begin{document}

\title{Localization and interactions in topological and non-topological bands in two dimensions}

\author{Akshay Krishna${}^1$, Matteo Ippoliti${}^2$, and R. N. Bhatt${}^1$}
\affiliation{${}^1$Department of Electrical Engineering and ${}^2$Department of Physics, Princeton University, Princeton NJ 08544, USA}

\begin{abstract}
A two-dimensional electron gas in a high magnetic field displays macroscopically degenerate Landau levels, which can be split into Hofstadter subbands by means of a weak periodic potential.
By carefully engineering such a potential, one can precisely tune the number, bandwidths, bandgaps and Chern character of these subbands.
This allows a detailed study of the interplay of disorder, interaction and topology in two dimensional systems.
We first explore the physics of disorder and single-particle localization in subbands derived from the lowest Landau level, that nevertheless may have a topological nature different from that of the entire lowest Landau level.
By projecting the Hamiltonian onto subbands of interest, we systematically explore the localization properties of single-particle eigenstates in the presence of quenched disorder.
We then introduce electron-electron interactions and investigate the fate of many-body localization in subbands of varying topological character.
\end{abstract}

\maketitle

\section{Introduction \label{sec:intro}}

The quantum Hall effect has been one of the most important and well studied physical phenomena in recent decades.
In the presence of a perpendicular magnetic field, the single-particle states of a two-dimensional electron gas reorganize themselves into highly degenerate Landau levels.
In the high field limit, where the cyclotron energy is much larger than all other energy scales in the problem, all the physics can be projected to the lowest Landau level (LLL) as inter Landau level mixing becomes negligible.
A disordered potential causes a broadening of the LLL and is well understood as the basis of the integer Quantum Hall transition \cite{Huckestein1995}.
In this case, there is a sub-thermodynamic number of extended states with Chern number $C = 1$ at the centre of the LLL \cite{Huo1992}.
On the other hand, in the presence of a periodic potential, the LLL splits into an intricate set of Hofstadter subbands \cite{Hofstadter1976}. 
The spectral properties and topological character of the subbands depend on the details of the periodic potential in a fine-tuned manner.

The addition of weak disorder to the Hofstadter problem causes the topologically non-trivial subbands to acquire critical energies, each with a diverging localization length  \cite{Tan1994, Huckestein1994, Koshino2005, Koshino2006}.
The scaling theory of localization predicts that all the single particle states in a two-dimensional subband will be localized if it does not have a topological character (i.e.\ its Chern number is zero).
When the strength of disorder is increased, the subbands broaden and critical energies corresponding to opposite Chern numbers annihilate each other in this process.
Finally, at a disorder much stronger than the periodic potential, the Hofstadter subbands are washed away and the LLL problem with a single critical energy is recovered.

The role of disorder in the presence of a periodic potential has been studied in continuum and lattice single-particle models as well as experimentally in the context of its implications on the Hall conductivity, electron localization and critical exponents \cite{Chalker1988, Aoki1996, Ishikawa1998, Yang1999, Huckestein1994, Bhaseen2000, Li2005, Slevin2009, Li2009, Obuse2010, Movilla2011, Lutken2014, Zhu2019}. 
However, the idea that the periodic potential can be carefully engineered to isolate both topological subbands with Chern numbers other than $+1$ (the LLL value), as well as topologically trivial subbands with zero Chern number, has not received as much attention in the studies of disordered systems.

In this paper, we first explore various ways of creating Hofstadter subbands with large bandgap-to-bandwidth ratios.
This enables us to safely neglect inter-subband-mixing and allows us to project all the physics to one or a few subbands, even when disorder and electron-electron interactions are present. 

It is well known that electron-electron interactions stabilize a gapped fractional quantum Hall phase at specific filling fractions of the LLL.
The addition of disorder causes a ground-state transition to an Anderson insulator \cite{Liu2016, Liu2017}.
However, the  highly excited states of the many-body spectrum do not show a corresponding localization transition \cite{Geraedts2017}.
This is argued to be a consequence of the diverging localization length in the single-particle spectrum, which delocalizes all many-body states in the presence of interaction \cite{Nandkishore2014A}.

Many-body localization (MBL) has been the subject of a considerable amount of work in the past decade~\cite{Nandkishore2015, Altman2015, Imbrie2017, Abanin2017, Alet2018, Parameswaran2018}.
The existence and stability of the MBL phase in one dimension is by now well established thanks to a combination of numerical and theoretical methods~\cite{Basko2006, Pal2010, Luitz2015, Wahl2017, Vosk2015, Potter2015, Dumitrescu2017, Thiery2018}, including a mathematical proof for the case of short-range interactions \cite{Imbrie2016, Imbrie2016B}.
However, many open questions remain, including the fate of the MBL phase in dimension greater than one~\cite{Lev2016, Inglis2016, Bertoli2018, Tomasi2018, Thomson2018, Wahl2019, Theveniaut2019}, the impact of long-range interactions\cite{Nandkishore2017}, and the importance of rare region effects on the stability of the MBL phase \cite{Nandkishore2014B, DeRoeck2017, Bordia2017, Potirniche2018}.

Our construction of nearly flat topological and non-topological bands in the LLL allows us to study the disordered and interacting problem in a projected Hilbert space in a two-dimensional continuum model.
It also enables us to decouple the effects of dimensionality and topology in destabilizing MBL in the LLL.

This paper is organized as follows. In Sec.\ \ref{sec:clean}, we set up the single-particle version of this problem and describe two methods for creating flat subbands. 
In Sec.\ \ref{sec:singleparticle}, we analyze the effects of disorder on single-particle flat-band models. We perform numerical exact diagonalization and use the inverse participation ratio of wavefunctions as a metric of localization. 
In Sec.\ \ref{sec:interaction}, we study the many-body problem in the presence of electron-electron interactions and use eigenvalue level statistics as well as memory of initial conditions under unitary dynamics as diagnostic tools for characterizing the ergodic-to-localized phase transition. 
We summarize our conclusions in Sec.~\ref{sec:discussion}.

This paper is an extension of our previous work \cite{Krishna2019}.
There, based on spectral statistics, we argued that there is evidence for an ergodicity breaking transition in topologically trivial LLL subbands in the one-dimensional limit.
Results for two dimensional scaling were not conclusive, showing a finite-size drift of the putative transition point.
The present work has two aims:
(i) to describe in detail the single-particle localization properties of the models studied, clarifying the genuine many-body nature of our previous results, and
(ii) to corroborate and extend those results.
In particular, these extensions include (i) a method for splitting the LLL into nearly-flat subbands of arbitrary Chern number, which enables the study of (de)localization in higher-Chern bands;
(ii) a more thorough two-dimensional finite-size scaling of one of our models; 
and (iii) an additional metric, beyond the level spacing statistic, to independently diagnose the breaking of ergodicity.


\section{The single-particle problem without disorder \label{sec:clean}}

In the high-field limit, the cyclotron energy and Zeeman splitting become arbitrarily large; as a result we can work with Hamiltonian of spin-polarized electrons
\begin{align}
H_{LLL} = \mathcal{P}_{LLL} V_{\text{1-body}} \mathcal{P}_{LLL}, \label{eq:hamiltonian1}
\end{align}
where $\mathcal{P}_{LLL}$ is the projector onto the LLL.
We consider a torus of dimensions $L_x$ and $L_y$, enclosing $N_\phi = \frac{L_x L_y}{2 \pi l_B^2}$ flux quanta, where the magnetic length $l_B = \sqrt{\hbar/eB}$.
We study two different examples of $V_{\text{1-body}}$.

\subsection{Lattice of point impurity scatterers \label{subsec:cleanlattice}}

\begin{figure}[ht!]
\centering
\includegraphics[trim={0 0.0cm 0 0.0cm}, clip, width=\columnwidth]{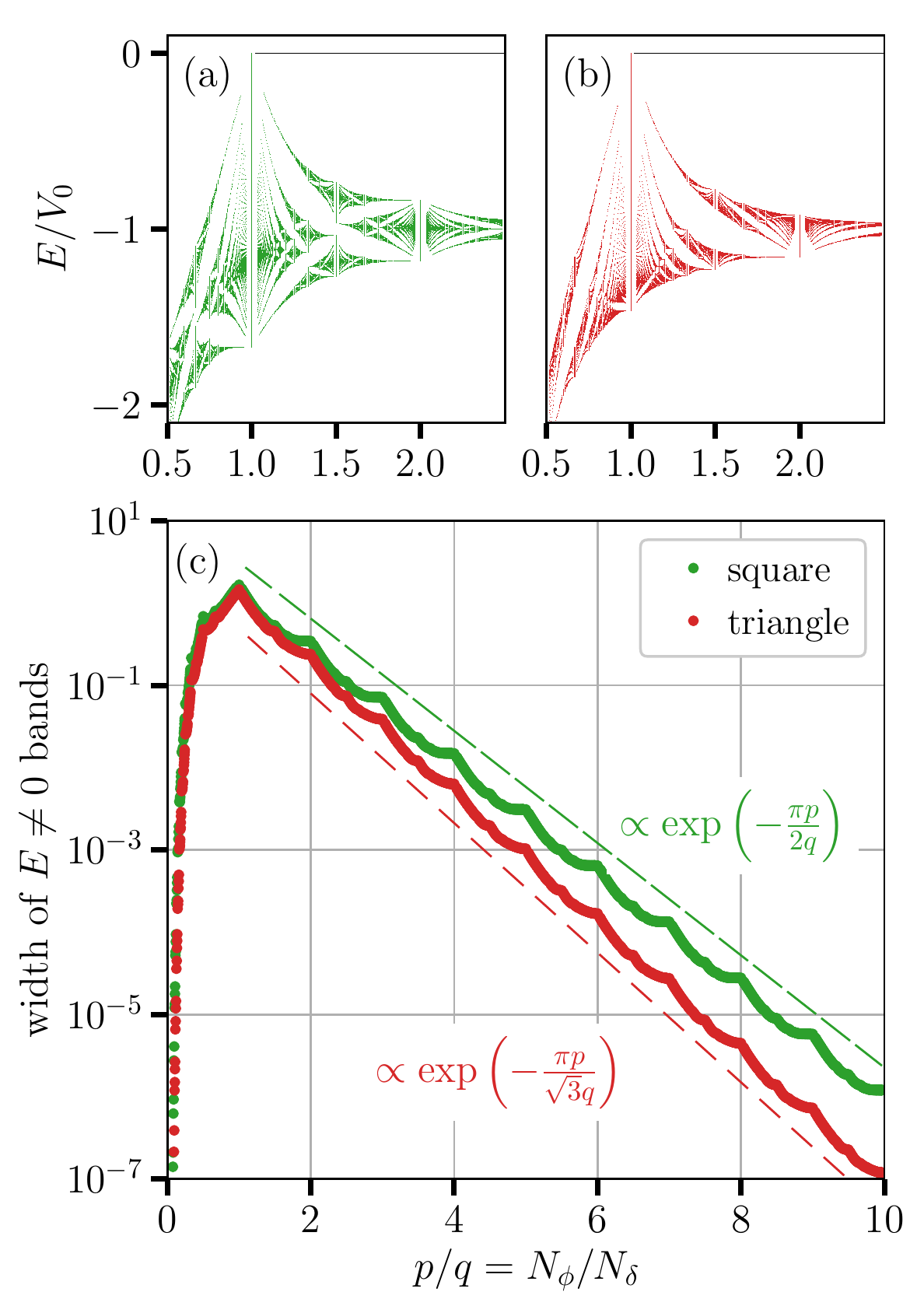}
\caption{(a) Hofstadter-like fractal band structure of a square lattice of point impurities as in Eq.\ \eqref{eq:deltapotential}, as a function of flux  per unit cell $p/q$. For $p/q > 1$, there is topological manifold at zero energy, and split-off bands with collective Chern number zero at negative energies. (b) Same as (a), but with a triangular lattice. (c) The total bandwidth of the split-off bands falls off exponentially with flux. The thin dashed lines are drawn as guides to the eye.
\label{fig:deltalattice}}
\end{figure}

The first model we consider is a lattice of point-like impurities, modeled by $\delta$ functions\cite{Ippoliti2018}:
\begin{align}
V_{\text{1-body}}(\mathbf{r}) = -V_0 \sum\limits_{n_1, n_2} \delta(\mathbf{r} - n_1 \mb{a}_1 - n_2 \mb{a}_2), \label{eq:deltapotential}
\end{align}
where $\mb{a}_1$ and $\mb{a}_2$ are the primitive lattice vectors.
The total number of delta functions $N_\delta$ is such that $\frac{N_\phi}{N_\delta} = \frac{p}{q}$, for coprime integers $p$ and $q$.
If $N_\delta < N_\phi$, we obtain $p - q$ degenerate bands at $E = 0$.
This manifold is topological with total Chern number $C=1$.
The remaining $q$ bands are non-topological and are centered around energy $-V_0$ (Fig.\ \ref{fig:deltalattice} (a,b)).

Our ultimate goal is to study the effect of disorder and interaction on the non-topological split-off bands derived from the LLL.
Thus we seek to pull out as many states as possible from the $E = 0$ band ($N_\delta \to N_\phi$).
However, this generically increases the bandwidth-to-bandgap ratio, as we will see next, so an optimum must be found. 

The width of the $E \neq 0$ split-off bands falls off exponentially with the amount of flux per delta function scatterer (Fig.\ \ref{fig:deltalattice}(c)).
For $\frac{p}{q} > 1$, this is nearly equal to the bandwidth-to-bandgap ratio.
In the dilute limit, each point impurity localizes one state of the form $\exp (- \frac{r^2}{2 l_B^2})$, having a Gaussian tail with length scale $l_B$.
In a tight-binding picture, the width of the band formed from these states depends on the overlap between nearest neighbor wavefunctions. 
This gives a bandwidth $\sim \exp \left( -\frac{r_n^2}{4 l_B^2} \right)$, where $r_n$ is the nearest neighbor distance.
The unit cell of a square lattice has area $r_n^2$, and encloses $p/q$ flux quanta.
Since an area of $2 \pi l_B^2$ encloses one flux quantum, the bandwidth for square lattices scales as $\exp \left( -\frac{r_n^2}{4 l_B^2} \right) = \exp \left( -\frac{\pi p}{2 q} \right)$.
For the same value of flux $p/q$, the triangular lattice gives a smaller bandwidth, scaling as $\exp \left( - \frac{\pi p}{\sqrt{3} q} \right)$.
In later sections, we specifically consider values of $p/q$ around 6, where the bandwidth-to-bandgap ratio is $\mathcal O(10^{-3})$.

\subsection{Smooth periodic potential \label{subsec:cleanpotential}}

More generally, we can take an arbitrary periodic potential $V(x,y) = V(x+a, y) = V(x, y+b)$ on a rectangular unit cell of size $a \times b$. 
In terms of the Fourier series \begin{align}
V_{\text{1-body}}(x, y) = \sum\limits_{m_x, m_y} v_{m_x, m_y} e^{i 2 \pi (m_x x/a + m_y y/b)} \;. \label{eq:Fourier}
\end{align}

If the unit cell encloses $\frac{p}{q}$ flux quanta, then the magnetic Brillouin zone is defined by quasimomenta $k_x$ and $k_y$ with $0 \leq k_x < \frac{2 \pi}{qa}$ and $0 \leq k_y < \frac{2 \pi}{b}$.
There are $p$ Hofstadter bands.

At a fixed $\mb{k}$, the band structure can be calculated by diagonalizing a $p \times p$ matrix, with elements $\mel{\psi_{\beta, \mb{k}}}{V_\text{1-body}}{\psi_{\beta', \mb{k}}}$.
The $\beta, \beta' \in \{ 0, 1, \cdots, p-1\}$ are band indices.
The Chern numbers $C$ of the subbands must obey the diophantine equation\cite{Thouless1982} $pC + qs = 1$, $s\in \mathbb Z$.

\begin{table}[ht!]
\begin{ruledtabular}
\begin{tabular}{c r r} 
 $({m_x, m_y})$ & $v_{m_x m_y} \times 10^{-3} $ & $\tilde{v}_{m_x m_y}$ \\ \hline
 $(1,0)$ & $4.16$ & $1895$ \\ 
 $(3,0)$ & $-35.80$ & $-30$\\ 
 $(1,1)$ & $-5.11$ & $-1062$ \\ 
 $(2,1)$ & $8.08$ & $159$ \\ 
 $(3,1)$ & $-38.98$ & $-15$ \\ 
 $(3,2)$ & $20.10$ & $0.7$\\
 $(3,3)$ & $20.66$ & $0.01$\\
 \end{tabular}
 \end{ruledtabular}
\caption{Values of Fourier components $v_{m_x, m_y}$ from Eq.\ \eqref{eq:Fourier}, that optimize the band flatness (width-to-gap ratio) at flux per unit cell $p/q = 2$. Square symmetry enforces $v_{m_x, m_y} = v_{m_x, -m_y} = v_{-m_x, m_y} = v_{m_y, m_x}$. The coefficients $\tilde{v}_{m_x,m_y} \equiv e^{-\frac{\pi}{4} (m_x^2+m_y^2)} v_{m_x,m_y}$ are rescaled by the form factor of the LLL. 
\label{tab:coefficients}}

\end{table}

We start by considering two fluxes per unit cell ($p=2$, $q=1$), which gives two subbands with Chern numbers $C=0$ and $C=+1$, respectively.
Our goal is to obtain widely separated bands with small dispersions $E_b$ and a large gap $E_g$, so that disorder $V_{\text{dis}}$ and interaction $V_{\text{int}}$ can lie in an intermediate range, $E_b \ll V_{\text{dis}}, V_{\text{int}} \ll E_g$.
We therefore optimize the periodic potential for maximal flatness $E_b / E_g \to 0$ in the space of square-symmetric potentials, with $|m_x|, |m_y| < 4$.
This results in the Fourier coefficients shown in Table \ref{tab:coefficients}.
The overall normalization (irrelevant to the $E_b/E_g$ ratio) is chosen to yield unit bandwidth $E_b = 1$.
Given the absence of terms with both $m_x$ and $m_y$ even, the resulting band structure has the symmetry $E_1(\mathbf{k}) = -E_2(\mathbf{k})$; the bands thus are identical except for their Chern numbers (see Appendix \ref{appendix:flatband} for details).
The listed Fourier coefficients yields a remarkably large bandgap-to-bandwidth ratio of $E_b/E_g \approx 8735$, allowing us to tune disorder and interaction over several orders of magnitude while safely neglecting inter-subband mixing.

\begin{figure}
\centering
\includegraphics[trim={0cm 0.0cm 0cm 0.0cm}, clip, width=0.96\columnwidth]{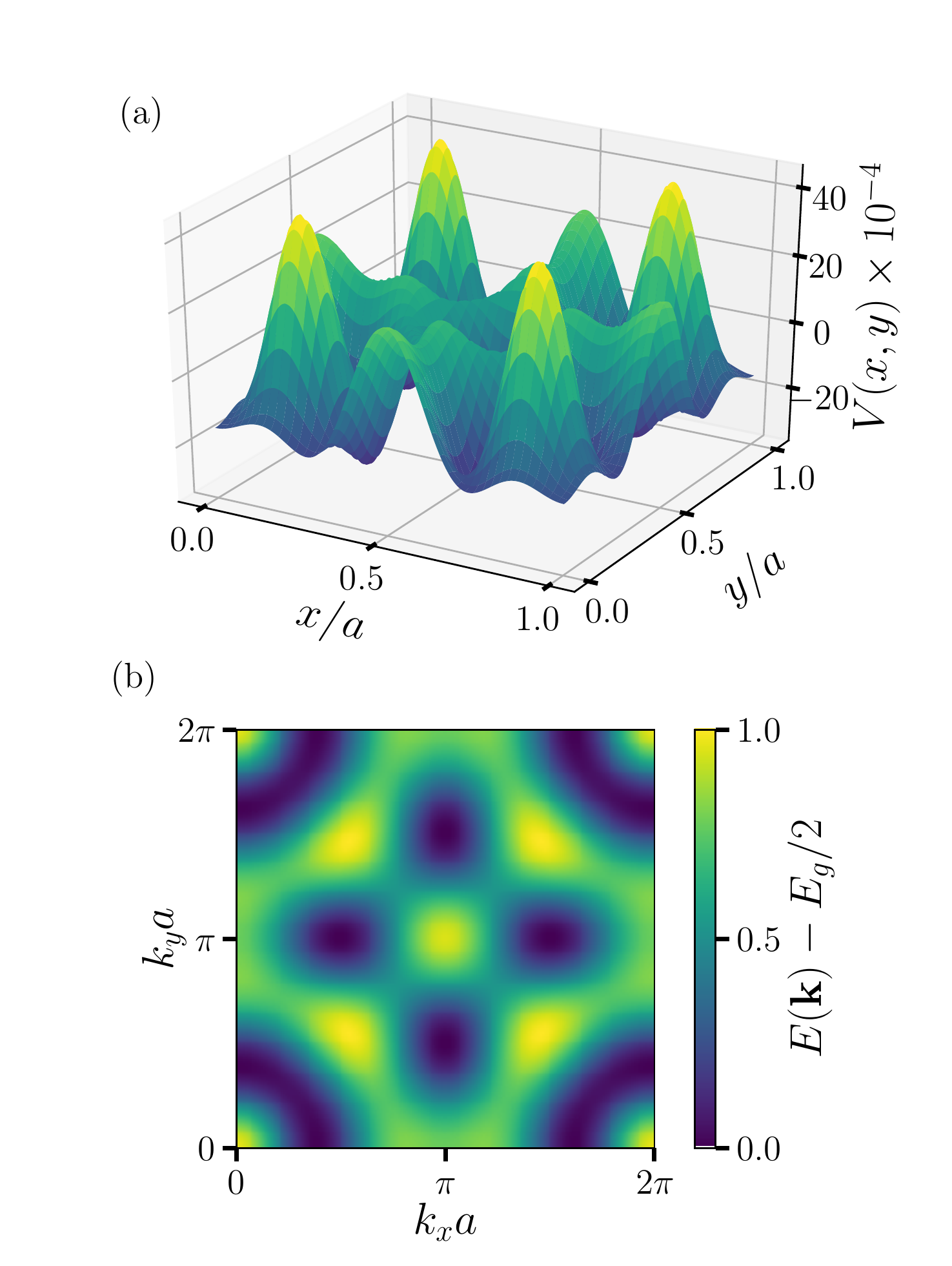}
\caption{(a) The potential $V(x, y)$ described by the Fourier coefficients in Table \ref{tab:coefficients} is plotted on one unit cell. (b) When a uniform magnetic field with two flux quanta per unit cell is applied, a band structure with extremely flat bands symmetric about $E=0$ is created. The upper $C=1$ band is plotted here. In units where the bandwidth is unity, the band gap $E_g \approx 8735$. The dispersion of the $C=0$ band is obtained by reflection.}
\label{fig:potential}
\end{figure}

We conclude by noting an interesting consequence of single particle Hamiltonian, which facilitates the engineering of flat bands with different Chern numbers. 
A pair of potentials $V$, $V^\prime$ may be constructed for flux-per-unit-cell values $\frac{p}{q}, \frac{p}{q'}$, where $q \neq q'$, such that $q-q'$ is a multiple of $p$, that yield \emph{exactly} the same band structure, $\{E_1(\mb k), \dots E_p(\mb k)\}$, \emph{except} for the Chern numbers of the bands.
This is accomplished by the rescaling
\begin{align}
v_{m_x,m_y}^\prime 
& = v_{m_x,m_y} e^{\frac{\pi}{2p}(q'-q) \left(m_x^2 \frac{b}{a} + m_y^2 \frac{a}{b} \right) } \;.
\label{eq:Vtrans}
\end{align}
Taking $q'>q$, the rescaling factor is rapidly divergent in $|\mb m|$, posing some constraints on the behavior of $v_{\mb m}$. However if $v$ has compact support (only a finite number of non-zero entries) then the transformation is always well defined.
As an example, we can take $\frac{p}{q'} = \frac{2}{3}$ and use the $v$ coefficients from Table~\ref{tab:coefficients} to obtain two nearly-flat bands with Chern numbers $C=+2$ and $-1$ respectively.
The proof of this construction is in Appendix \ref{appendix:flatband}.
In the rest of the paper, we study the effects of disorder and interaction on these identically dispersive nearly flat Chern subbands with $C = -1, 0, 1$ and $2$.


\section{Disorder and single-particle localization \label{sec:singleparticle}}

In the absence of disorder, both models introduced in Sec.~\ref{sec:clean} feature extended Bloch eigenfunctions.
In the presence of quenched disorder, the electronic eigenstates acquire a finite localization length $\xi$.
In the following, we calculate $\xi$ as a function of disorder strength for the two models introduced in Sec.~\ref{sec:clean}.
For each eigenstate, calculated by exact diagonalization, we estimate the localization length $\xi = \frac{1}{\sqrt{2 \pi P_2}}$, where the inverse participation ratio $P_2 \equiv \int d^2 \mathbf{r}\ |\psi(\mathbf{r})|^4$.
This definition ensures that a purely exponentially localized wavefunction $\psi(\mathbf{r}) \sim e^{-r/\xi}$ has a localization length $\xi$.

The energy-resolved ensemble-averaged localization lengths $\langle \xi(E) \rangle$ are computed for different finite sizes.
$\langle \xi(E) \rangle $ is usually peaked close to the center of the band, and falls off near the tails of the band.
In the following, we use the maximal localization length $\bar{\xi} \equiv \max\limits_E  \langle \xi(E) \rangle$ as a measure of the localization of the wavefunctions in a band.
Another effect of quenched disorder is an increase in the energy bandwidth $E_b$ of a band.
In this section, we quantify the impact of disorder on localization in the LLL subbands using these two measures $\bar{\xi}$ and $E_b$ for two different models.

\subsection{Disordered point impurities \label{subsec:lattice}}

We start with the lattice of point impurities  and randomize the strengths and positions of scatterers by replacing Eq.\ \eqref{eq:deltapotential} with \begin{align}
V_{\text{1-body}}(\mathbf{r}) = - \sum\limits_{n=1}^{N_\delta} V_{n} \delta(\mathbf{r} - \mathbf{r}_n). \label{eq:deltapotential_dis}
\end{align}

The $V_{n}$ are independently and identically distributed uniform random variables in $[1-W, 1 + W]$.
In this paper, we consider $0 < W < 1$.
The positions of the scatterers $\mathbf{r}_n$ are randomly distributed on the torus, with a circular exclusion zone around each scatterer of area $2 \pi l_B^2 \frac{N_\phi}{N_\delta} \rho$. 
Placing impurities randomly allows us to circumvent the constraints imposed by lattice-based models.
Similar to the lattice case, if $N_\delta < N_\phi$, there is a manifold of $N_\phi - N_\delta$ degenerate states at zero energy with total Chern number $C=1$. 

The width of the remaining $N_\delta$ split-off states is controlled by the disorder, which has two independent components.
The randomness in the scatterers' strengths is controlled by $W$, and randomness in their positions is controlled by the density parameter $0 < \rho < \frac{\pi}{2 \sqrt{3}} = 0.907$.
The upper bound for $\rho$ comes from a triangular lattice, which is the closest possible packing in two dimensions.
For large $\rho$, the distribution of scatterers becomes more regular (the maximum value indeed forces the configuration to be a triangular lattice with no randomness left). 
At the opposite end, $\rho=0$ corresponds to maximal randomness and allows two scatterers to sit arbitrarily close to each other, thus entirely closing the band gap 
(if two scatterers sit at the same exact position, then $N_\delta \mapsto N_\delta-1$, and the split-off band loses one state).

\begin{figure}[ht!]
\centering
\includegraphics[width=0.97\columnwidth]{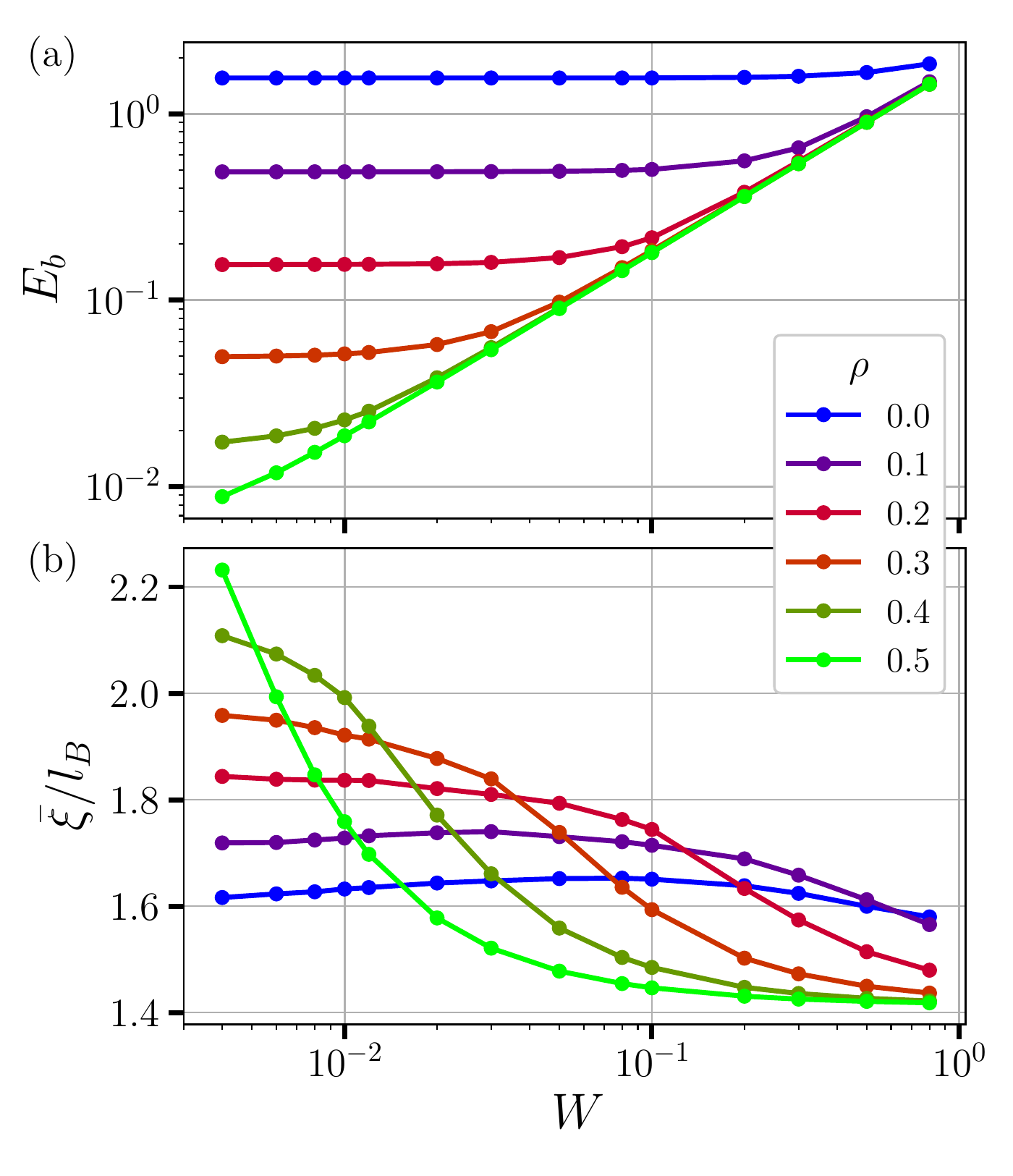}
\caption{(a) The bandwidth $E_b$ of the split-off states for the model described in Eq.\ \eqref{eq:deltapotential_dis} is plotted as a function of amplitude disorder $W$ and density parameter $\rho$. The bandwidth $E_b$ is defined as the energy interval within which 90\% of the split-off states lie. (b) The maximum average localization length $\bar{\xi}$ of the band is calculated using the inverse participation ratio as described in the text. We set $N_\phi = 6 N_\delta$ and perform exact diagonalization of 200 realizations of disorder at each parameter value. The system size is $N_\phi = 3000$ (torus dimension $\approx 137 l_B$).}
\label{fig:random_delta_1d}
\end{figure}

In Fig.\ \ref{fig:random_delta_1d}, we explore the parameter space of positional and amplitude disorder in this model.
We set the number of scatterers $N_\delta = N_\phi/6$, so that there are effectively 6 flux quanta per point impurity. 
For this density of scatterers, the bandwidth of the split-off states is $<10^{-3}$ in the disorder-free case (see Fig.\ \ref{fig:deltalattice}).
In general, the bandwidth $E_b$ of the split-off states increases with disorder.
At a fixed value of positional disorder $\rho$, there is a regime where the bandwidth $E_b$ and the  maximum localization length of the band $\bar{\xi}$ are largely independent of amplitude disorder $W$.
As $W$ is increased, there is a transition to a regime where the amplitude disorder $W$ becomes more relevant. In this regime $\bar{\xi}$ decreases with increasing $W$, and $E_b$ increases linearly with $W$.
The localization length at the center of the split-off band, $\bar{\xi}$, is very small ($\ll 137 l_B$, the linear dimension of the torus studied).

For the purpose of the many-body problem, which we will discuss in Sec.~\ref{sec:interaction}, we are interested in a regime where the bandwidth is small (so the projection is justified), yet the states respond strongly to changes in disorder (so an ergodic-to-localized transition is plausible).
For this reason, we fix $\rho = 0.4$ (close to midway between completely disordered and jammed scatterers) and vary $W$ in the vicinity of $10^{-2}$.

\subsection{Periodic potential with correlated disorder \label{subsec:potential}}

To localize states in the flat Hofstadter subbands generated by the continuum periodic potential of Eq.\ (\ref{eq:Fourier}), we introduce a short-range correlated disorder represented by a potential $V_{dis}(\mathbf{r})$ with 
\begin{equation}
\langle V_{\rm dis}(\mb r) V_{\rm dis}(0) \rangle = W^2 \sigma^{-2} e^{-r^2/2\sigma^2} \;. \label{eq:corr_dis}
\end{equation}
Here $\langle \cdots \rangle$ denotes averaging over realizations, $\sigma$ is a correlation length and $W$ quantifies the strength of disorder.
Setting $\sigma = 0$ yields uncorrelated Gaussian white noise.

\begin{figure}[ht!]
\centering
\includegraphics[width=1.00\columnwidth]{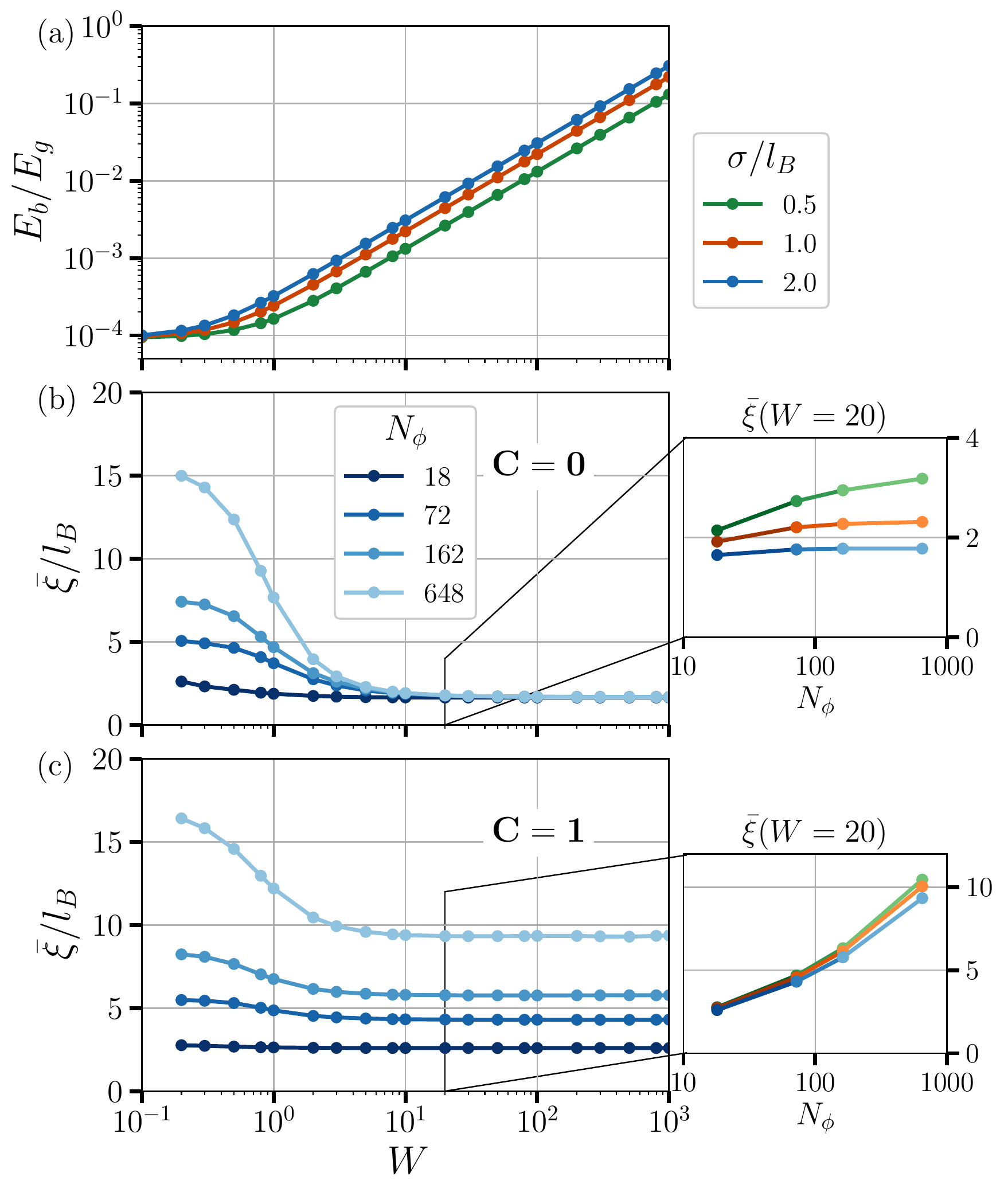}
\caption{(a) Similar to Fig.\ \ref{fig:random_delta_1d} (a), the bandwidth-to-bandgap ratio $E_b/E_g$ of the Hofstadter subbands of the continuum periodic potential of Fig.\ \ref{fig:potential} is plotted as a function of the strength $W$ of the correlated disorder potential (Eq.\ \eqref{eq:corr_dis}) for three different values of disorder correlation length $\sigma$. Each of the two bands has the same bandwidth $E_b$ within numerical accuracy. (b) The maximum average localization length $\bar{\xi}$ of the $C=0$ subband is plotted as a function of disorder $W$ and system size $N_\phi$. The disorder correlation length is fixed at $\sigma = 2 l_B$. The inset on the right shows the saturation of $\bar{\xi}$ with system size at $W=20$, for different values of disorder correlation length $\sigma$. (c) Same as (b), but for the $C=1$ subband. Unlike the previous case, the localization length diverges with increasing system size.}
\label{fig:loc1body}
\end{figure}

As in the previous case, we explore the single particle localization properties as a function of the disorder parameters.
In this case, the much larger bandwidth-to-bandgap ratio of the disorder-free problem allows us to vary $W$ over a large dynamic range without mixing the two bands.
We seek a parameter regime where the localization length is rather small even for moderate disorder and small system sizes.

\begin{figure}[ht!]
\centering
\includegraphics[width=1.00\columnwidth]{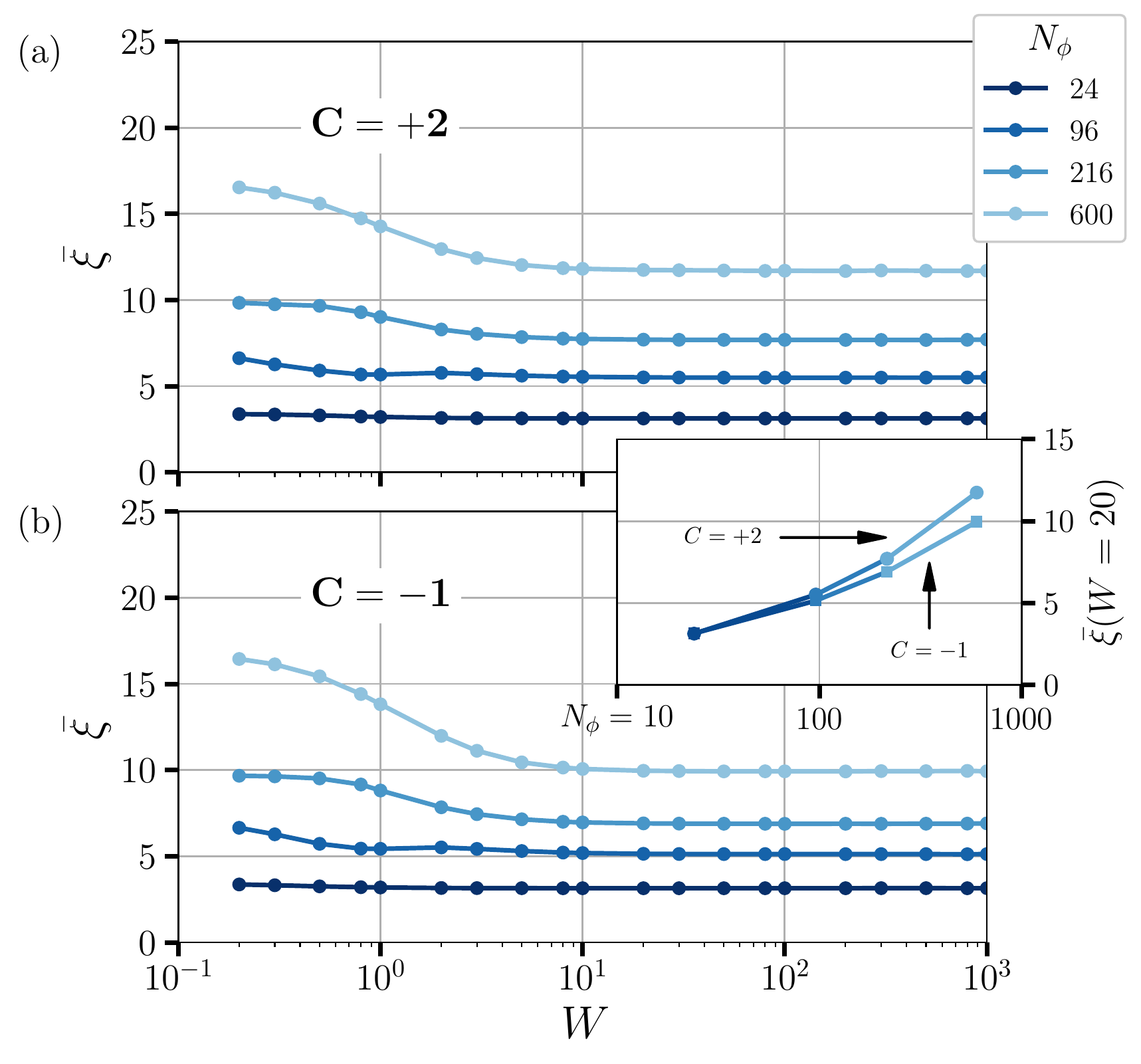}
\caption{The maximum average localization lengths $\bar{\xi}$ of the $C=+2$ and $C=-1$ subbands are plotted as a function of the correlated disorder strength $W$ for the model with $2/3$ flux quanta per unit cell. The correlation length of the disorder potential $\sigma = 2 l_B$. The inset shows the scaling of $\bar{\xi}$ as a function of system size at fixed disorder $W = 20$.}
\label{fig:loc1body_p2_q3}
\end{figure}

In Fig.\ \ref{fig:loc1body}, we study the bandwidth and localization length of states in the $C=0$ and $C=1$ bands as a function of disorder strength and correlation length.
Disorder makes the bands broaden to an equal extent, but we are able to attain a large amplitude of disorder $W \approx 1000$ without closing the gap.
For the $C=0$ band, the localization length saturates to a small value ($\sim 2.5 l_B$) for large enough disorder $W \gtrsim 10$ for all system sizes.
We also observe (inset) that using a larger correlation length $\sigma$ makes the localization length of the $C=0$ band saturate much faster with system size.
This is especially useful towards the many-body problem, where system sizes accessible by numerical diagonalization are by necessity very small.
For this reason, in Sec.~\ref{sec:interaction} we shall fix $\sigma = 2l_B$ when considering this model.
On the contrary, for the $C=1$ band, the localization length grows with system size without bound, indicating the presence of a divergent localization length in the thermodynamic limit, i.e. a critical energy much like in the whole LLL.
The scaling of $\bar{\xi}$ is expected to be described by the non-trivial multifractal scaling exponents of the integer quantum Hall effect \cite{Huckestein1995, Evers2008B}.

The periodic potential with coefficients as in Table \ref{tab:coefficients} is rescaled by Eq.\ \eqref{eq:Vtrans} and unit cell is simultaneously resized to admit $2/3$ flux quanta per unit cell.
This yields a band structure exactly as for the previous case (see Fig.\ \ref{fig:potential}), but with subbands having Chern numbers $+2$ and $-1$ respectively.
When correlated disorder is added to this model, we see in Fig.\ \ref{fig:loc1body_p2_q3} that both subbands have a diverging localization lengths $\bar{\xi}$.
The $C=+2$ subband has two critical energies in the middle of the band, and shows stronger delocalization \cite{Yang1996, Yang1999, Wan2001}.

In the many-body case, we would therefore expect the $C=+2$ subband to resist MBL more strongly than the $C=\pm 1$ subbands.



\section{Localization with interactions \label{sec:interaction}}

In Sec.~\ref{sec:clean} we presented two methods to engineer nearly-flat subbands with varying topological character from the LLL, and in Sec.~\ref{sec:singleparticle} we added disorder to both and studied their single-particle localization properties.
We next turn on electron-electron interactions and turn our attention to the possibility of many-body localization in these models.

While the model based on a smooth periodic potential has the advantage that it allows us to project the disordered potential onto very flat bands of arbitrary Chern number, the periodicity of the potential places a severe restriction on the sizes we can access numerically in the interacting case.
This constraint is absent in the model based on point impurities, which allows greater flexibility in choosing system sizes at the cost of (i) lack of tunability of Chern numbers, (ii) a strong asymmetry between the $C=0$ and $C=+1$ subbands, and (iii) worse band flatness ratio.
The two models thus feature distinct strengths and weaknesses as platforms for the study of localization with interactions. 

\subsection{Method \label{subsec:method}}

We consider the Hamiltonian
 \begin{align}
H_{LLL} = \mathcal{P}_{sb} \mathcal{P}_{LLL} \left[ V_{\text{1-body}} + V_{\text{int}} \right] \mathcal{P}_{LLL} \mathcal{P}_{sb}, \label{eq:mbhamiltonian}
\end{align}
where $V_{\text{int}}$ is an interaction term and $\mathcal{P}_{sb}$ further projects the LLL Hamiltonian onto a single subband of desired Chern character obtained through the methods discussed in previous sections.
With $N_e$ electrons, the filling fraction $\nu = N_e/N_o$, where $N_o$ is the number of single particle orbitals in the projected subband.

We choose $V_{\text{int}}$ to be a Haldane $V_1$ pseudopotential interaction $V_{\text{int}}(\mathbf{k}) = V_c (1 - k^2 l_B^2) e^{-k^2 l_B^2/2}$.
There are now two independent parameters -- the interaction strength $V_c$, and the disorder strength characterized by $W$ (and $\rho$, in the point-impurity model) as described in the previous section.

We investigate the onset of a possible many-body localization transition by two methods -- eigenvalue statistics and infinite-time persistence of an initial charge density imbalance. 

We compute the many-body eigenvalues $\{ E_n\}$ and eigenvectors $\{ \ket{\phi_n} \}$ of the Hamiltonian in Eq.\ \eqref{eq:mbhamiltonian} via numerical exact diagonalization on rectangular tori of dimensions $L_x \times L_y$.
Then we compute the eigenvalue spacing ratio \begin{align}
r_n = \frac{\min \left(E_n - E_{n-1}, E_{n+1} - E_{n} \right)}{\max \left(E_n - E_{n-1}, E_{n+1} - E_{n} \right)}.
\end{align}
The ensemble-averaged mean value of this statistic $\langle r \rangle$ is a useful diagnostic of localization-delocalization transitions \cite{Oganesyan2007, Pal2010, Luitz2015}.
In a delocalized (thermal) phase, the eigenvalue statistics are governed by the Gaussian unitary ensemble (GUE), characteristic of an ergodic system with broken time-reversal symmetry.
In this case, $\langle r \rangle \approx 0.5996$ \cite{Atas2013}.
In a disordered (localized) phase, the eigenvalue spacing distribution is Poissonian and $\langle r \rangle = 2 \ln 2 - 1 \approx 0.3862$.


A popular experimental method to probe many-body localization and prethermalization is the time evolution of an initial density imbalance in the system \cite{Schreiber2015, Smith2016, Choi2016, Neyenhuis2017, Luschen2017, Luschen2017B, Kohlert2019}.
In the ergodic phase, unitary time evolution from any initial state should scramble the system completely, so the initial imbalance should vanish at long times.
On the other hand, in the many-body localized phase, memory of initial conditions is retained to arbitrarily long times under unitary evolution, so a finite residual imbalance should be observed even after infinite time.

In order to probe this effect numerically, we consider the relaxation of an initial charge density imbalance, modeled by the traceless Hermitian operator
\begin{align}
M = \int \mathrm{d}r \ c^\dagger_{\mb{r}} c_{\mb{r}} \cos (2 \pi x / L_x)\;.
\end{align}
We initialize a mixed state close to infinite temperature with density matrix $\rho_{0} = \frac{1 + \epsilon M}{\dimension}$, 
where $\epsilon$ is a small positive coefficient and $\dimension$ is the Hilbert space dimension.
The initial amplitude of charge density imbalance is\begin{align}
\langle M_0 \rangle &= \Tr \left[ \rho_{0} M \right] = \frac{\epsilon}{\dimension} \Tr M^2 \\
&=  \frac{\epsilon}{\dimension} \sum\limits_m \expval{\phi_m | M^2 | \phi_m}.
\end{align}
This means the system initially has higher charge density near $x\simeq 0$ than it does at the opposite side of the torus, near $x\simeq L_x/2$. 
Unitary evolution will relax this imbalance in the ergodic phase but not completely in the MBL phase.

The amplitude of the charge density imbalance at times $t>0$ is given in terms of the time-evolved density matrix
\begin{equation}
\rho_t = e^{-iHt} \rho_0 e^{iHt} = \sum_{m,n} e^{i (E_n-E_m)t} | \phi_m \rangle \langle \phi_m | \rho_0 | \phi_n \rangle \langle \phi_n | \;.
\end{equation}
At long times $t \gg \hbar / \delta E$ (where $\delta E$ is the typical many-body energy spacing), the off-diagonal density matrix elements accumulate phases that time-average to zero (we assume the disorder prevents any degeneracies).
In this limit, using the $x_\infty$ notation as shorthand for $\lim_{T\to\infty} \frac{1}{T} \int_0^T dt\ x(t)$, we have
\begin{equation}
\rho_{\infty} = \sum_m ( \langle \phi_m | \rho_0 | \phi_m \rangle )  | \phi_m \rangle  \langle \phi_m |
\end{equation}
and therefore
 \begin{align}
\langle M_\infty \rangle &= \Tr \left[ \rho_{\infty} M \right] 
= \frac{\epsilon}{\dimension} \sum\limits_m \left[ \expval{\phi_m | M| \phi_m} \right]^2.
\end{align} 

The remnant charge imbalance is the ratio $\langle M_\infty / M_0 \rangle$.
It lies between zero and one and quantifies the extent to which the initial charge density modulation is `remembered' at infinite time.
It provides us with a useful metric to complement the level statistics to diagnose the lack of ergodicity and thus the possibility of a many-body localization transition.

\begin{table}[ht!]
\begin{ruledtabular}
\centering
\begin{tabularx}{0.4\columnwidth}{c c c c}
 $N_\phi$ & $N_\delta$ & $N_e$ & $L_x/l_B = L_y/l_B$ \\ \hline
54 & 9 & 3 & $6 \sqrt{3 \pi}$ \\
72 & 12 & 4 & $12 \sqrt{\pi} $ \\ 
90 & 15 &  5 & $6 \sqrt{5 \pi} $ \\ 
108 & 18 &  6 & $6 \sqrt{6 \pi} $ \\ 
\end{tabularx}
\end{ruledtabular}
\caption{
Summary of system sizes studied in the point impurity model, as described in Sec.\ \ref{subsec:lattice}. We exactly diagonalize four different system sizes, each with square aspect ratio, at a filling $\nu = N_e / N_\delta =  1/3$. The number of scatterers is $N_\delta = N_\phi/6$.
\label{tab:runs_delta}}
\end{table}

In the following we apply this method to the two models previously discussed.
We first study the disordered distribution of point impurities.
This allows us to study 2-D finite size scaling in the non-topological $C=0$ subband with a fixed filling, which we set to $\nu=1/3$.
Next, for the smooth periodic potential of Sec.\ \ref{subsec:cleanpotential}, we solve the problem on rectangular tori of dimensions $L_x \times L_y$ at various fillings.
This system gives us access to $C=0$, $C=\pm 1$ and $C=2$ subbands, depending on the flux per unit cell.
The Hilbert space dimension grows exponentially with the system size, limiting the maximum size.
Further, we must ensure that the torus admits an integer number of unit cells with periodic boundary conditions.

\begin{figure}[ht!]
\centering
\includegraphics[trim={0 0.0cm 0 0.0cm}, clip, width=\columnwidth]{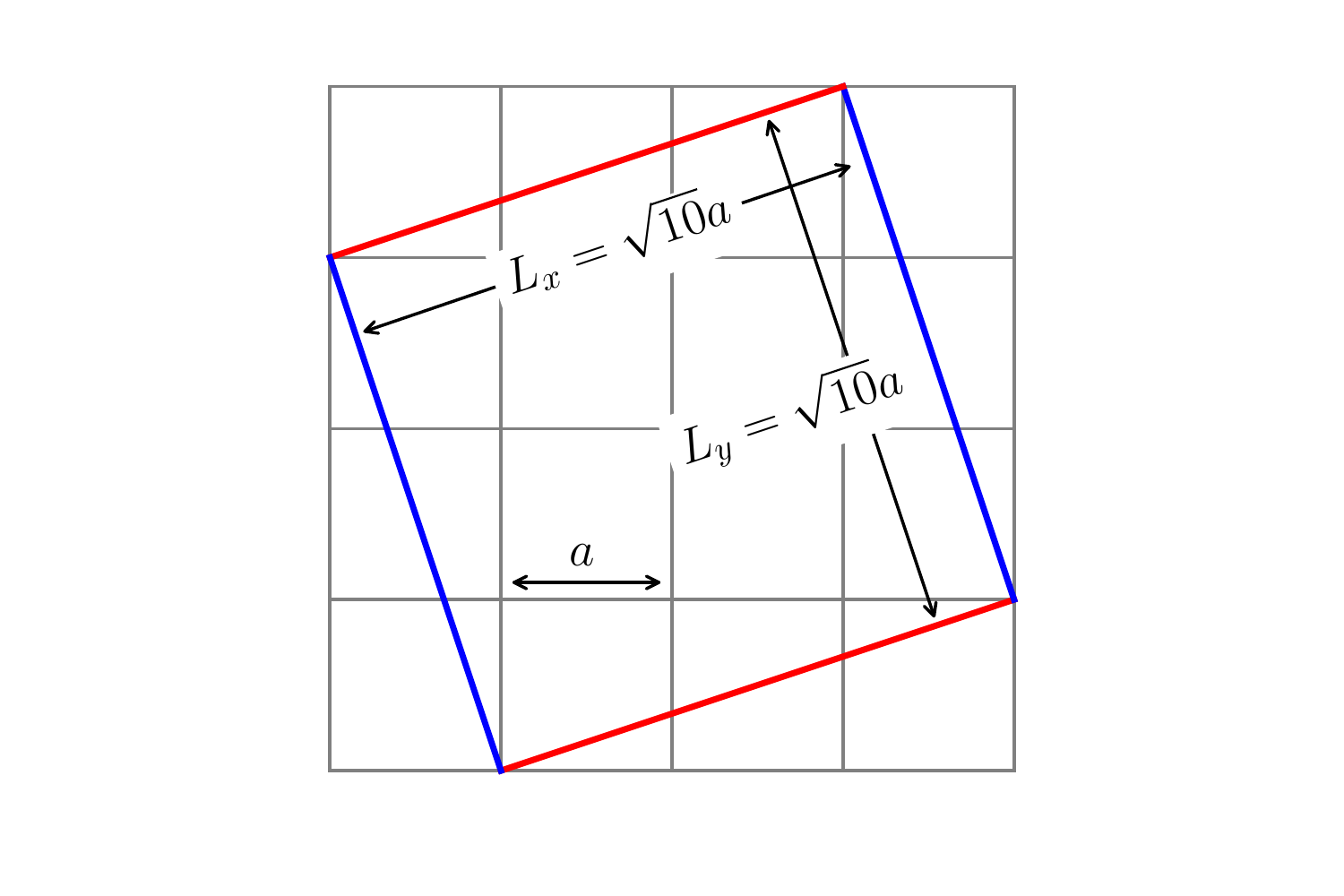}
\caption{The torus of size $L_x \times L_y$ can be rotated with respect to the underlying periodic potential. This allows us to obtain tori with linear dimensions that are an irrational multiple of the lattice constant. In this example, the torus encloses an area of $10 a^2$. Periodic boundary conditions are imposed on parallel edges as indicated.
\label{fig:rotatedlattice}}
\end{figure}

\begin{table}
\begin{ruledtabular}
\begin{tabularx}{\columnwidth}{c  c  c  c  c}
 $N_\phi$ & $\frac{L_x}{a} \times \frac{L_y}{a}$ & $\nu_{\text{min}}$ & $\nu_{\text{max}}$ & $\Delta \nu$\\ \hline
16 & $2\sqrt{2} \times 2\sqrt{2}$ & $0.25$ & $0.5$ & $0.125$ \\
18 & $3 \times 3$ & $0.222$ & $0.444$ & $0.1111$  \\
20 & $\sqrt{10} \times \sqrt{10}$ & $0.2$ & $0.5$ & $0.1$  \\
26 & $\sqrt{13} \times \sqrt{13}$ & $0.154$ & $0.462$ & $0.0769$\\
32 & $4 \times 4$ & $0.125$ & $0.5$ & $0.0625$ \\
34 & $\sqrt{17} \times \sqrt{17}$ & $0.118$ & $0.471$ & $0.0588$ \\
36 & $3\sqrt{2} \times 3\sqrt{2}$ & $0.111$ & $0.5$ & $0.0556$ \\ \hline
24 & $3 \times 4$ & $0.167$ & $0.5$ & $0.0833$  \\
30 & $3 \times 5$ & $0.133$ & $0.467$ & $0.0667$  \\
36 & $3 \times 6$ & $0.167$ & $0.5$ & $0.0556$  \\
\end{tabularx}
\end{ruledtabular}
\caption{Summary of system sizes studied in the smooth potential model of Sec.~\ref{subsec:potential} with $p/q=2$.
The unit cell has side is $a = 2 \sqrt{\pi} l_B \approx 3.54 l_B$. 
For each value of $N_\phi$, the accessible fillings $\nu = 2 N_e/N_\phi$ lie between $\nu_\text{min}$ and $\nu_\text{max}$, spaced by $\Delta \nu$.
We perform calculations for seven different sizes for 2-D scaling on square tori, and three different sizes for quasi 1-D scaling on rectangular tori.
\label{tab:runs_p2_q1}}
\end{table}

If we keep the orientation of the torus aligned with that of the lattice, so that both $L_x$ and $L_y$ are multiples of the lattice constant $a$, we are severely limited in the sizes we can access.
This limitation is partially alleviated by rotating the torus with respect to the lattice, as illustrated in Fig.\ \ref{fig:rotatedlattice}.
In the case of $p/q = 2$ flux quanta per unit cell, the system sizes we can access are listed in Table \ref{tab:runs_p2_q1}.

We perform two kinds of analysis.
First, at a fixed filling of $\nu = 1/3$ and $L_x = 3a$, we compare the results at different values of $L_y$.
This is equivalent to quasi 1-dimensional scaling.
Second, we attempt a 2-dimensional scaling by comparing the results for square tori $L_x = L_y$ of different sizes.
In the latter case, no one filling is available at every size, therefore we interpolate the data to estimate the scaling behavior at a fixed filling $\nu = 1/3$.

\begin{table}[ht!]
\begin{ruledtabular}
\begin{tabularx}{\columnwidth}{c c c c c}
 $N_\phi$ & $\frac{L_x}{a} \times \frac{L_y}{a}$ & $\nu_{\text{min}}$ & $\nu_{\text{max}}$ & $\Delta \nu$\\ \hline
12 & $3 \sqrt{2} \times 3\sqrt{2}$ & $0.333$ & $0.5$ & $0.167$\\
24 & $6 \times 6$ & $0.167$ & $0.5$ & $0.0833$\\
30 & $3\sqrt{5} \times 3\sqrt{5}$ & $0.133$ & $0.467$ & $0.0667$ \\
\end{tabularx}
\end{ruledtabular}
\caption{Same as Table \ref{tab:runs_p2_q1}, but for $p/q = 2/3$ flux quanta per unit cell. The period of the smooth potential is $a = 2\sqrt{\pi/3}l_B \approx 2.05 l_B$.
In this case, we perform calculations only on square tori.
\label{tab:runs_p2_q3}}
\end{table}

Finally, for the case where each unit cell encloses $p/q  = 2/3$ flux quanta (where the subbands have Chern numbers $C=+2$ and $C=-1$), the system sizes are listed in Table \ref{tab:runs_p2_q3}.

\subsection{Results \label{subsec:results}}

\begin{figure}[ht!]
\centering
\includegraphics[trim={0 0.0cm 0 0.0cm}, clip, width=\columnwidth]{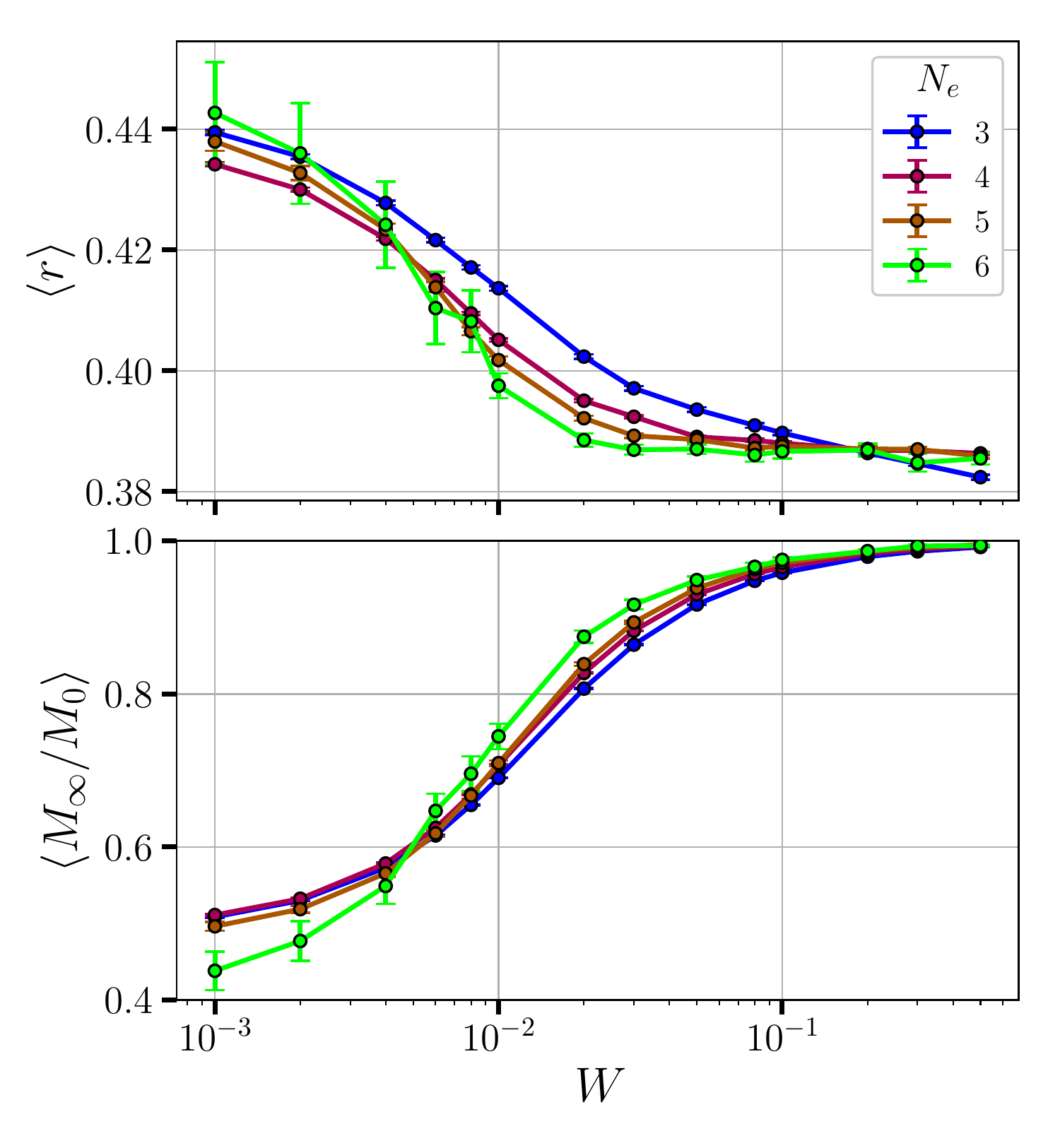}
\caption{Two-dimensional scaling of the $\langle r \rangle$ statistic (above) and remnant charge imbalance $\langle M_\infty / M_0 \rangle$ (below) as a function of disorder $W$ for electrons at a filling fraction $\nu = 1/3$ in the $C = 0$ subband of the disordered point impurity model.
Details of system sizes are given in Table \ref{tab:runs_delta}. 
\label{fig:r_delta_lattice}}
\end{figure}

In Fig.\ \ref{fig:r_delta_lattice}, we show the $\langle r \rangle$ statistic and the persistence of charge imbalance $\langle M_\infty / M_0 \rangle$ as a function of disorder strength for the $C=0$ of the disordered distribution of point impurities.
We fix the interaction strength $V_c=1$, and the non-interacting bandwidth $E_b \lesssim 0.1$ as in Fig.\ \ref{fig:random_delta_1d} (a).
 
\begin{figure}[ht!]
\centering
\begin{tabular}{cc}
  \includegraphics[trim={1cm 1.4cm 0.7cm 0.6cm}, clip, width=0.235\textwidth ]{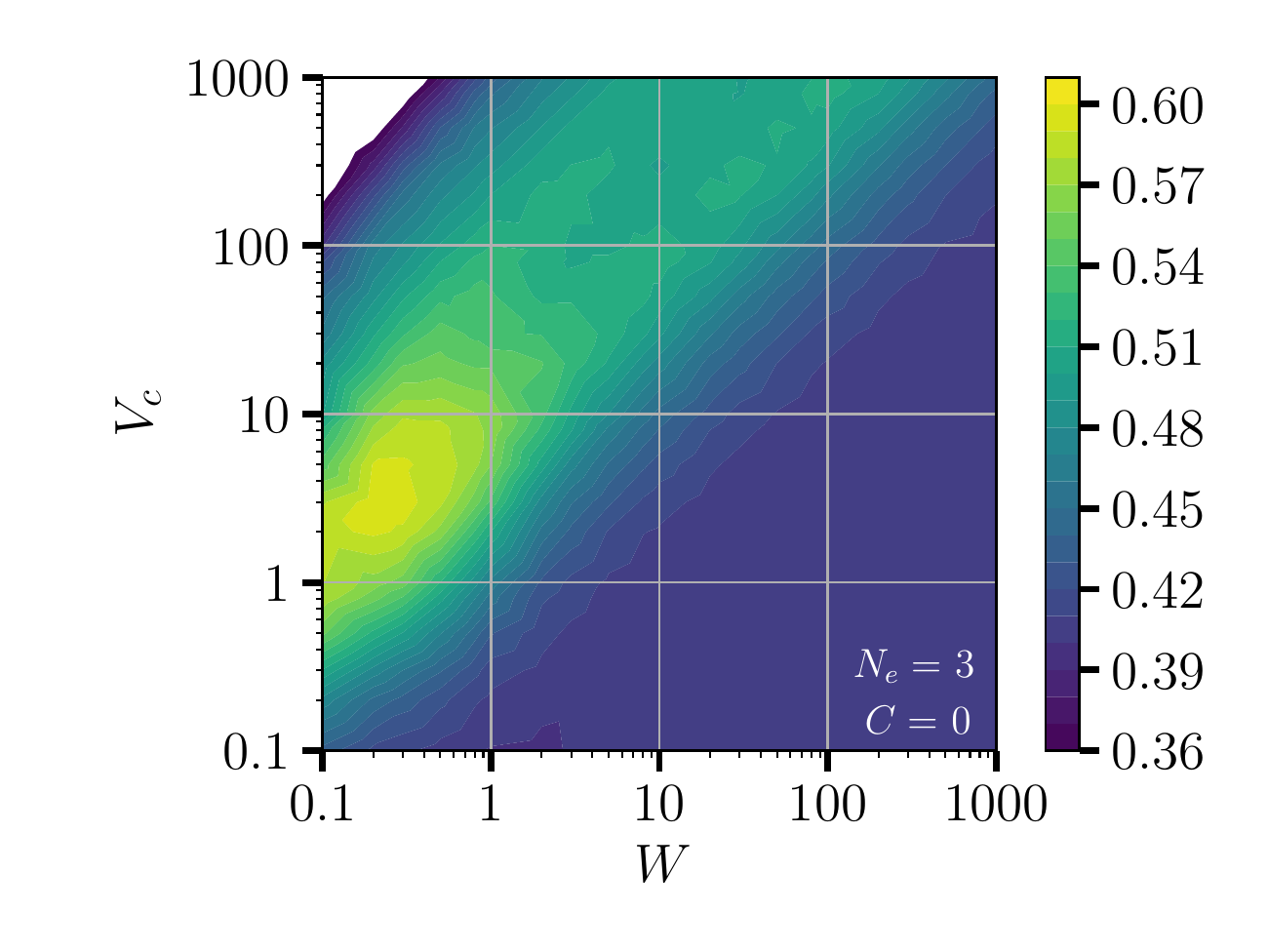} &   \includegraphics[trim={1cm 1.4cm 0.7cm 0.6cm}, clip, width=0.235\textwidth ]{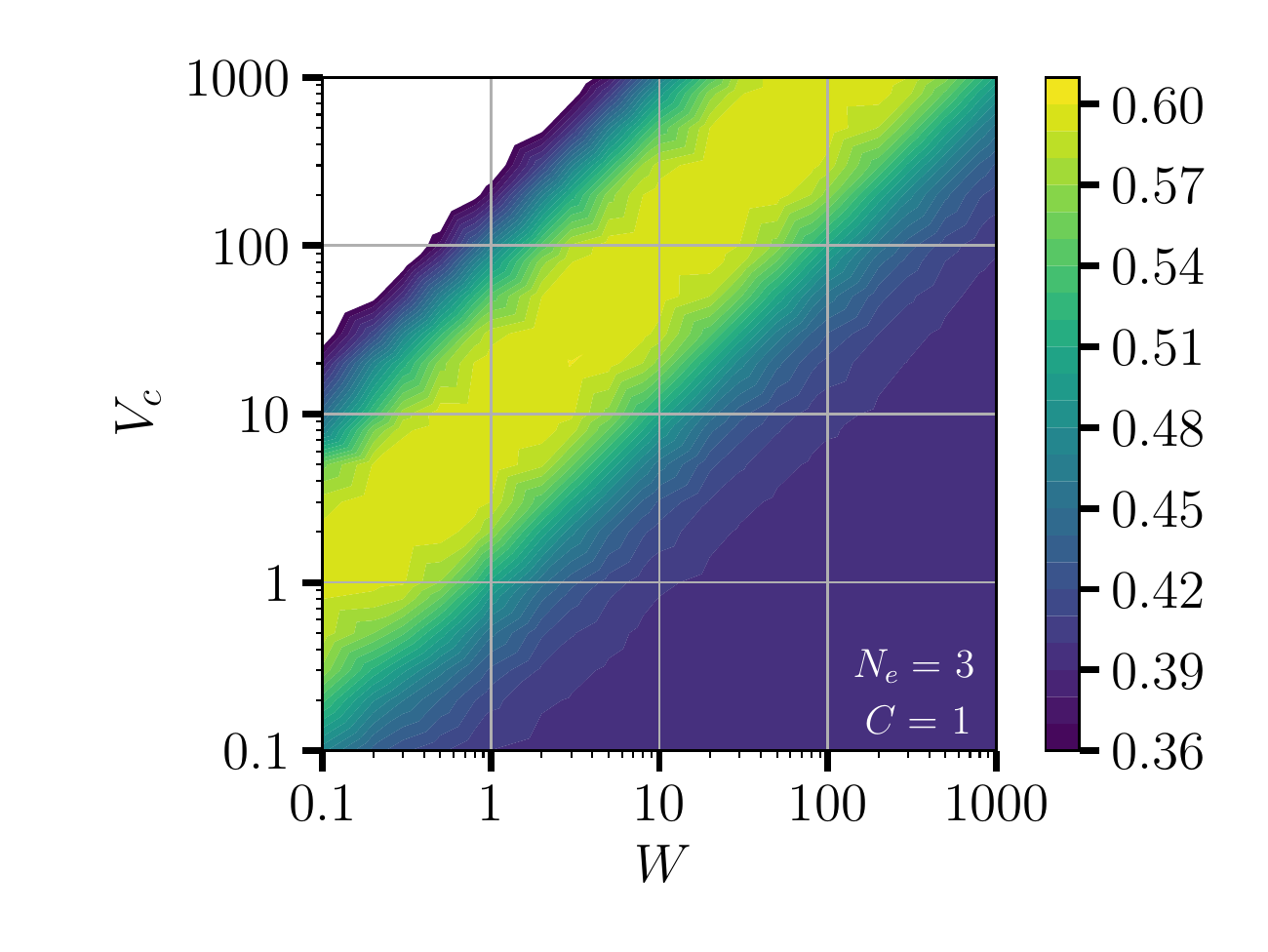} \\
  \includegraphics[trim={1cm 1.4cm 0.7cm 0.6cm}, clip, width=0.235\textwidth ]{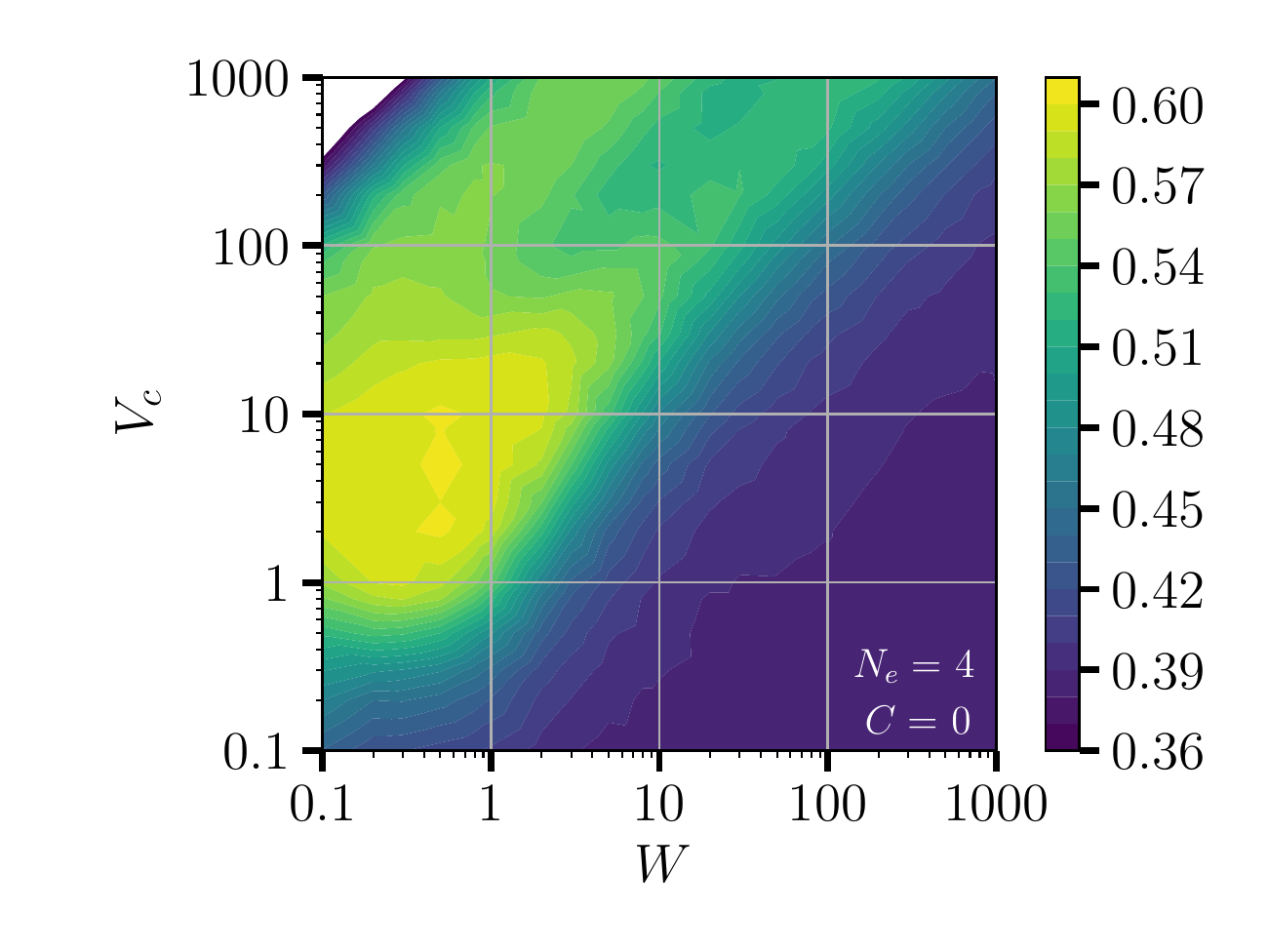} &   \includegraphics[trim={1cm 1.4cm 0.7cm 0.6cm}, clip, width=0.235\textwidth ]{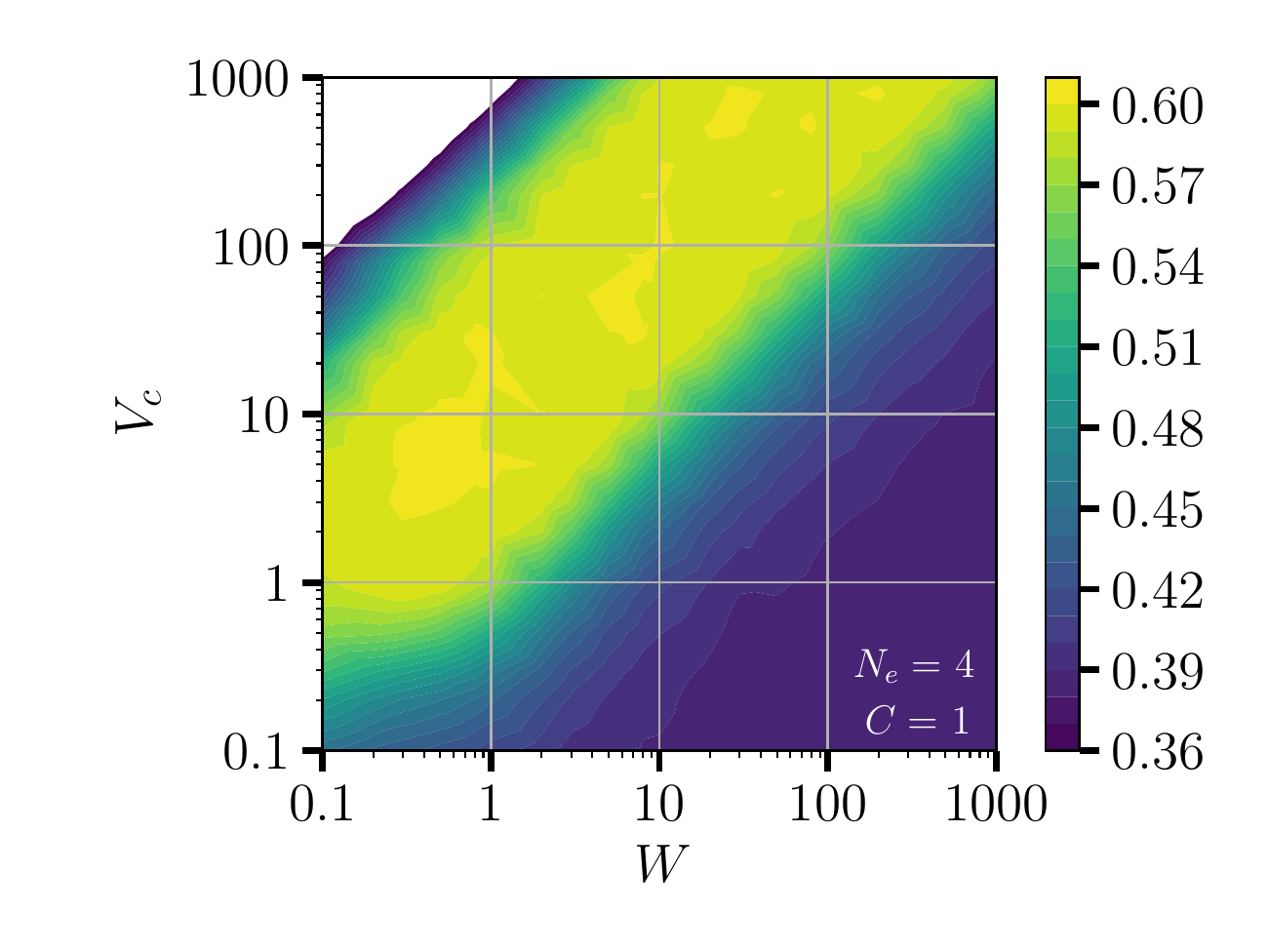} \\
  \includegraphics[trim={1cm 1.4cm 0.7cm 0.6cm}, clip, width=0.235\textwidth ]{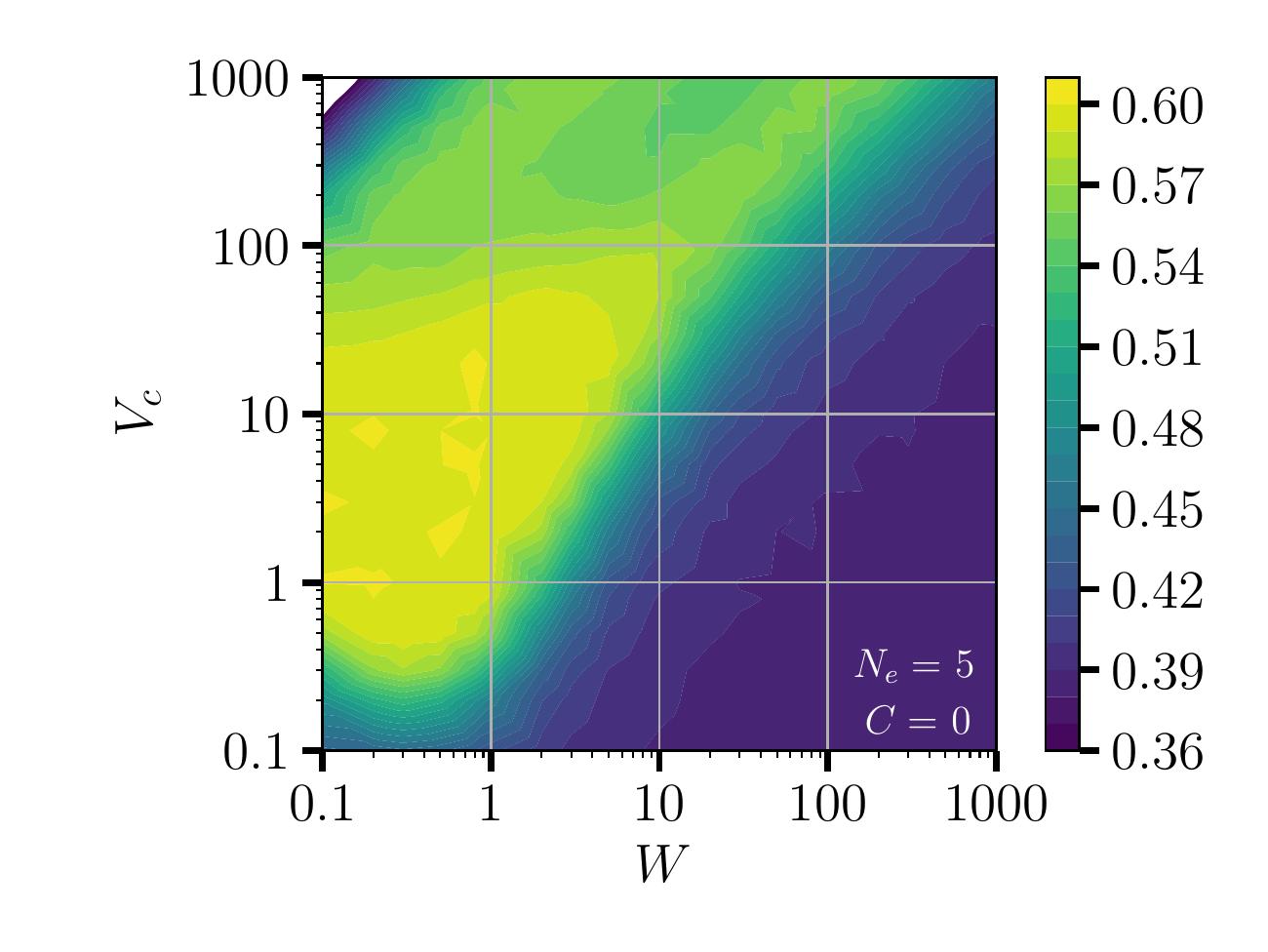} &   \includegraphics[trim={1cm 1.4cm 0.7cm 0.6cm}, clip, width=0.235\textwidth ]{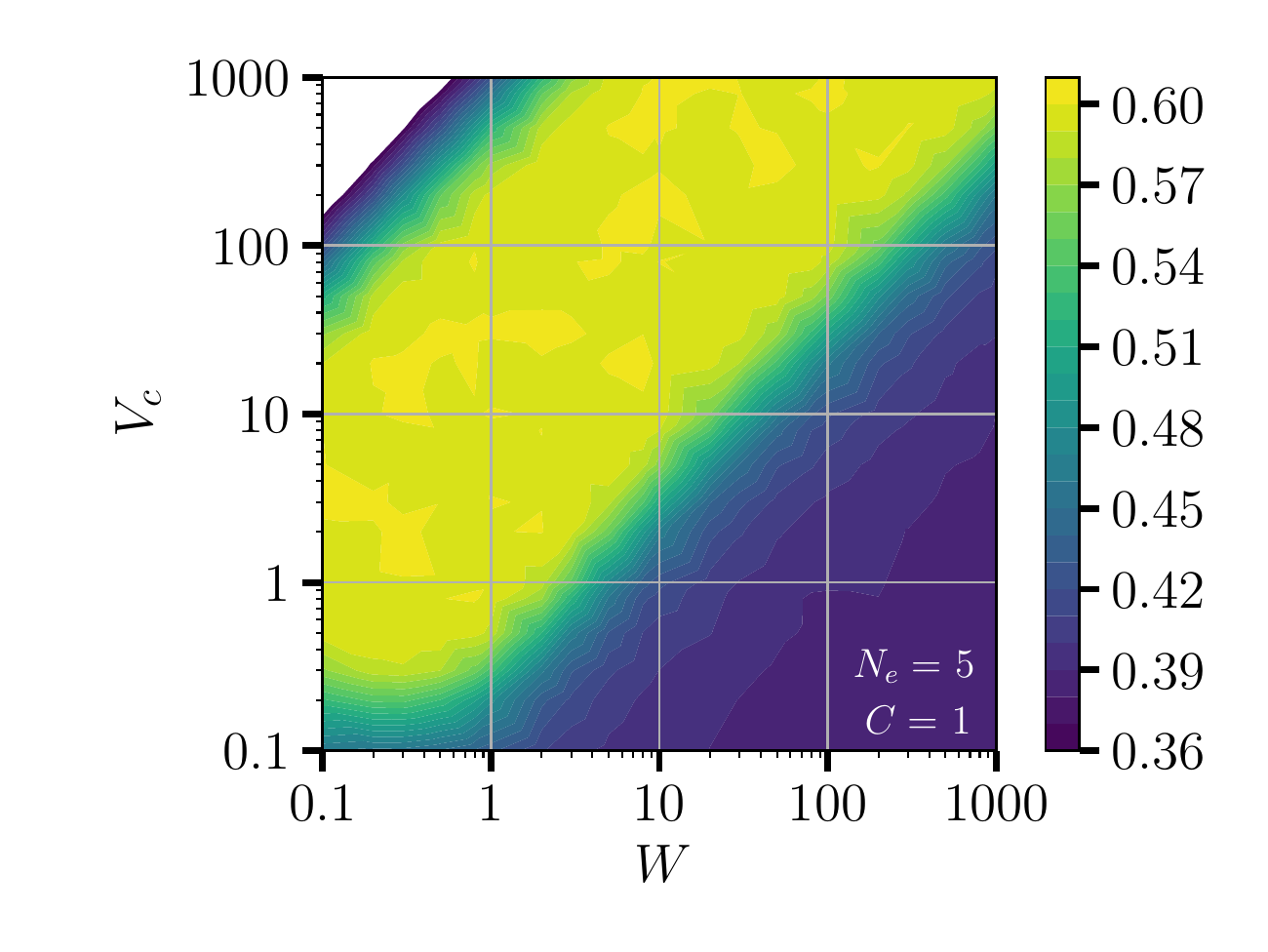}\\
  \includegraphics[trim={1cm 0.2cm 0.7cm 0.6cm}, clip, width=0.235\textwidth ]{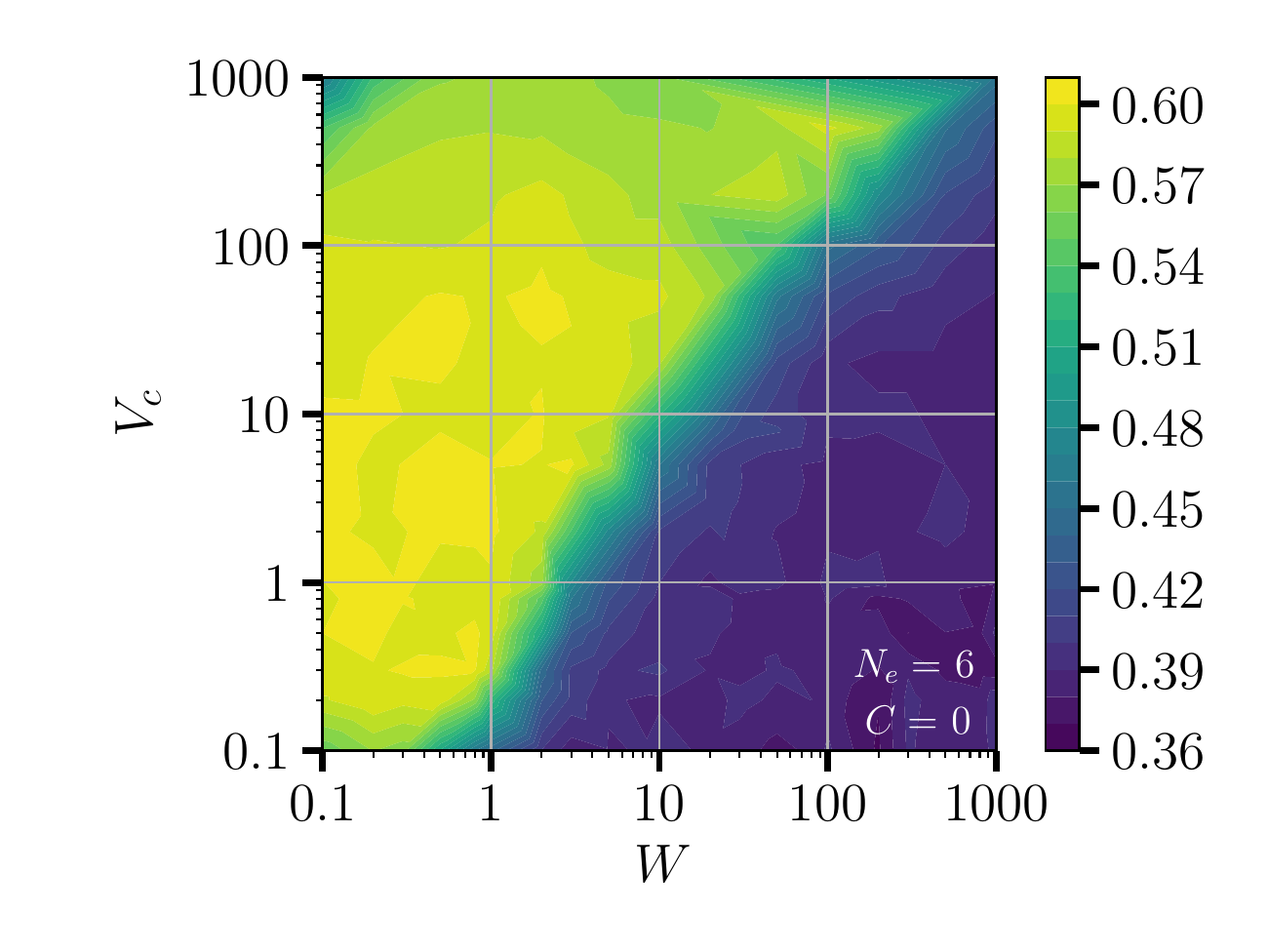} &   \includegraphics[trim={1cm 0.2cm 0.7cm 0.6cm}, clip, width=0.235\textwidth ]{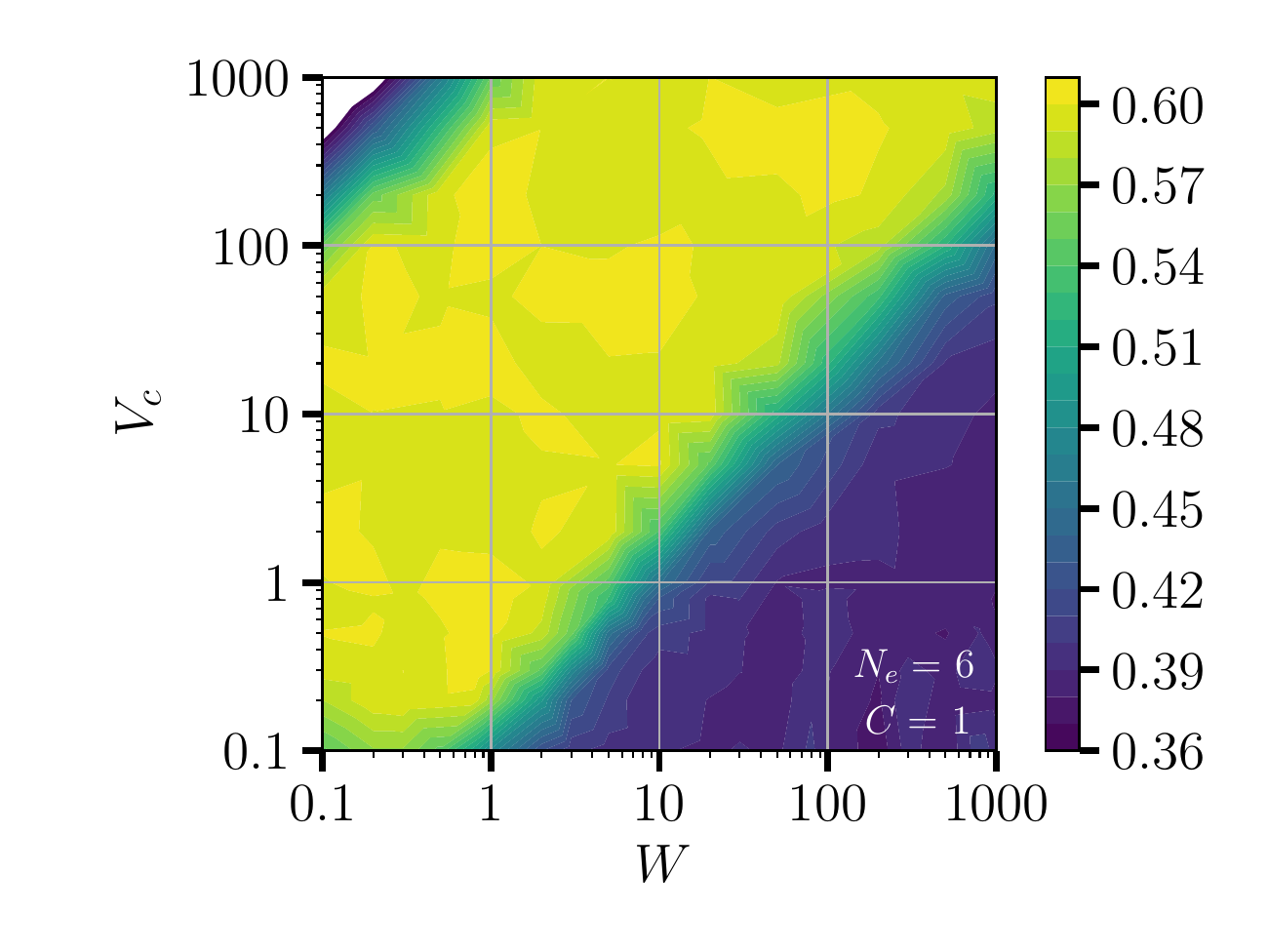}
\end{tabular}
\caption{The $\langle r \rangle$ statistic is plotted as a function of the interaction strength $V_c$ and disorder strength $W$ for a filling of $1/3$ for the $C=0$ (left) and $C=1$ (right) bands, for four different system sizes ($N_\phi = 18, 24, 30, 36$), increasing from top to bottom.
The $C=1$ band shows an increasing tendency to delocalize as the system size is increased.
Approximately 100 different disorder realizations are averaged over at $N_e=6$ and around 10000 at $N_e = 6$.
One linear dimension of the torus is kept fixed ($L_x$) while the other is increased.
\label{fig:r_W_Vc}
}
\end{figure}

\begin{figure*}[ht!]
\centering
\includegraphics[trim={0 0.0cm 0 0.0cm}, clip, width=\textwidth]{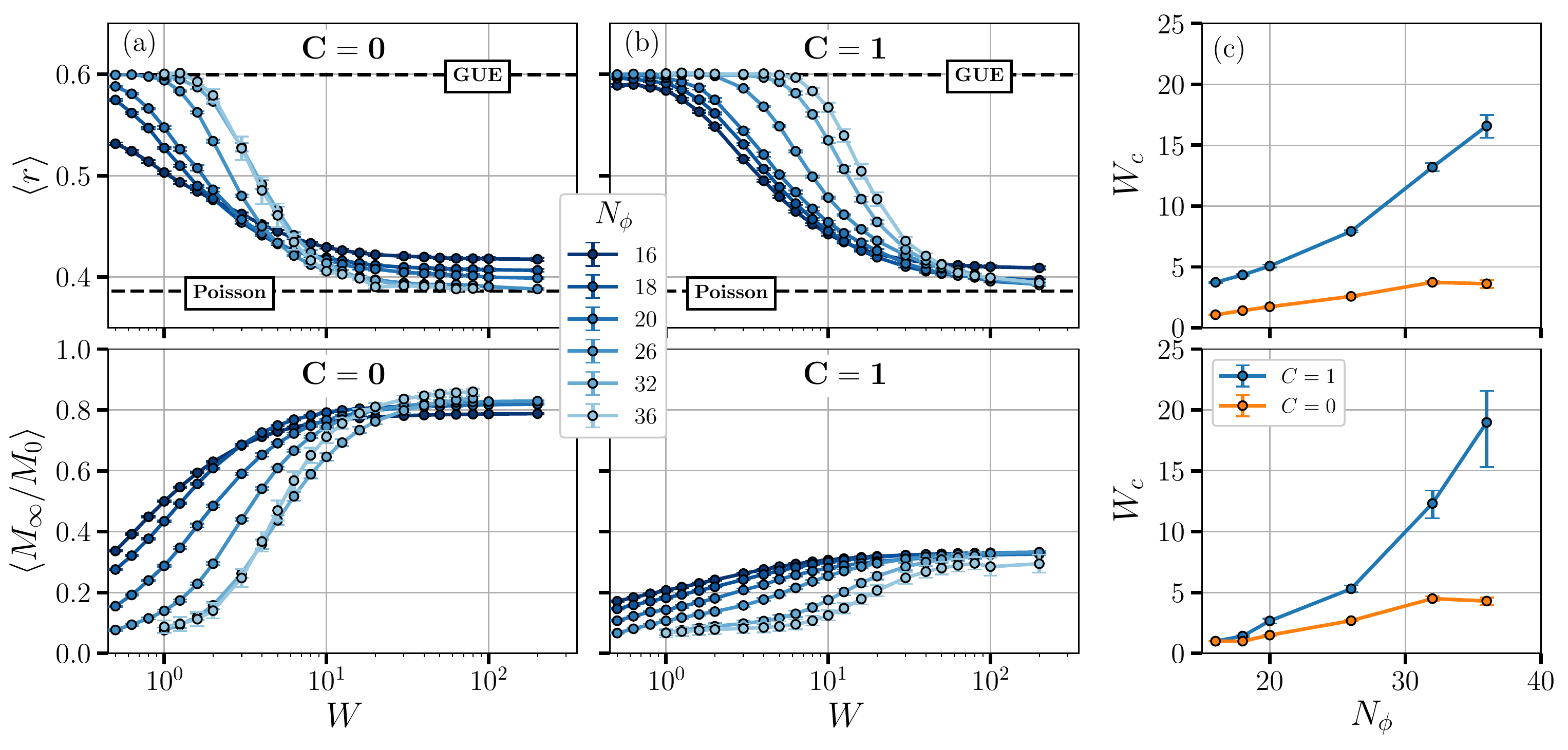}
\caption{The $\langle r \rangle$ statistic (above) and remnant charge imbalance $\langle M_\infty / M_0 \rangle$ (below) are plotted against disorder both for the $C=0$ (a) and $C=1$ subbands (b) at a filling of $\nu = 1/3$. For values of $N_\phi$ not divisible by 3, the curves are obtained by interpolating from the nearest available rational fraction. The $\langle r \rangle$ statistic of the $C=0$ subband attains the localized Poisson value at much smaller disorder than the $C=1$ subband. The value of $\langle M_\infty / M_0 \rangle$ is also much larger for the $C=0$ subband, and also appears to flow towards a step function as system size is increased.
In (c), we plot the critical disorder $W_c$ as a function of system size $N_\phi$.
We define $W_c$ to be the value of W at which $\langle r \rangle = 0.5$ (roughly halfway between Poisson and GUE) in the upper panel.
In the lower panel, we define $W_c$ to be the value of W at which $\langle M_\infty / M_0 \rangle$ attains roughly half its saturation value.
For the $C=0$ subband, we define $\langle M_\infty / M_0 \rangle (W_c) = 0.4$, and for the $C=1$ subband, we define $\langle M_\infty / M_0 \rangle (W_c) = 0.2$.
\label{fig:r_p2_q1_2d}}
\end{figure*}

This model was shown to have clear evidence of a crossing of the $\langle r \rangle$ statistic between Poisson and random matrix behavior at a filling $\nu = 1/2$ around $W = 10^{-2}$ for quasi 1-D scaling \cite{Krishna2019}.
Two dimensional scaling also showed such crossing, but it was unclear whether that behavior persisted in the thermodynamic limit or whether it slowly drifted to infinite disorder.
Here, we study a different filling of $\nu = 1/3$.
The $\langle r \rangle$ statistic decreases with disorder $W$, as expected, and also shows a crossing between different sizes around $W \approx 10^{-2}$.
The persistence of charge imbalance $\langle M_\infty / M_0 \rangle$ also shows a clear monotonically increasing trend as a function of disorder.
We observe a crossing in the curves at $W \approx 10^{-2}$, consistent with the eigenvalue statistics.
The lack of access to larger system sizes precludes us from making a conclusive statement about whether this behavior is characteristic of a finite temperature transition in the thermodynamic limit.
It is possible that there is a slow drift of the crossing with system size, indicating the instability of true MBL in this system in the two-dimensional limit.
The many-body localization transition is a high temperature phenomenon, and is not expected to depend on the properties of the ground state.
That our results for $\nu = 1/2$ and $\nu = 1/3$ are so similar is consistent with this expectation.

Next, we turn to the smooth potential of Sec.\ \ref{subsec:cleanpotential}, and perform quasi 1-D scaling at filling $\nu = 1/3$. In Fig.\ \ref{fig:r_W_Vc}, we show the $\langle r \rangle$ statistic for both the $C=0$ and $C=1$ subbands.
It is evident that although disorder and interaction provide two independent energy scales $W$ and $V_c$ respectively, only their ratio $W/V_c$ seems to matter in controlling the ergodic-to-MBL transition, as long as $W$ and $V_c$ are both sufficiently larger than the bandwidth (which we set to 1).
In the rest of the paper, we fix the interaction strength at $V_c = 8$, which is a factor of $10^3$ smaller than the inter-subband gap, and large enough to be in the regime where results depend solely on $W/V_c$.
Both subbands show an increasing tendency to delocalize as the system size is increased. However, the critical disorder strength $W_c$ at which $\langle r \rangle$ is midway between Poisson and GUE statistics seems to appproach a finite value for the $C=0$ subband, while that for the $C=1$ subband seems to diverge with system size.

In Fig.\ \ref{fig:r_p2_q1_2d}, we plot the eigenvalue statistic $\langle r \rangle$, and the persistence of charge imbalance $\langle M_\infty / M_0 \rangle$ as a function of disorder strength for 2-D scaling on square tori.
Values are obtained by interpolating the curves as a function of filling for different sizes to obtain curves at $\nu = 1/3$ (see Appendix \ref{appendix:nu_interp} for details).
In the $C=0$ subband, it is evident that there is a signature of a finite disorder transition.
At small $W$, the remnant charge imbalance is close to zero, indicating a thermal phase in which memory of initial conditions is washed away completely.
At large $W$, the remnant charge imbalance is non-zero.
The transition between the two regimes becomes sharper as the system size is increased, indicating that the phenomenon is likely to persist in the thermodynamic limit.
This behavior is mirrored in the eigenvalue $\langle r \rangle$ statistic, which smoothly interpolates between the GUE value at small disorder and the Poisson value at large disorder with a crossing very close to the Poisson value, in line with previous works \cite{Luitz2015}.

\begin{figure*}[ht!]
\centering
\includegraphics[trim={0 0.0cm 0 0.0cm}, clip, width=\textwidth]{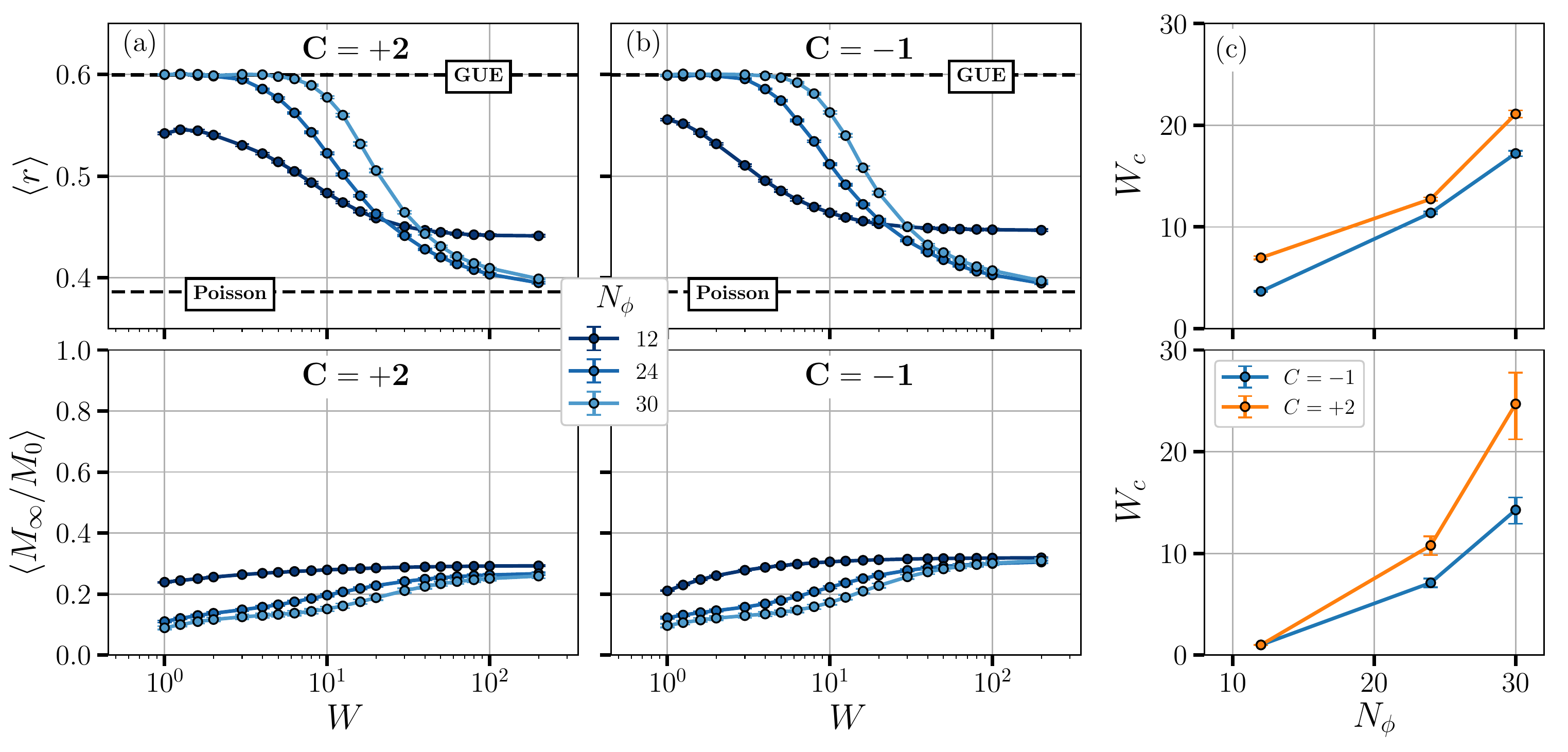}
\caption{Same as Fig.\ \ref{fig:r_p2_q1_2d}, but for $p/q = 2/3$ flux quanta per unit cell. The two subbands have Chern numbers $+2$ and $-1$ respectively.
Both the $\langle r \rangle$ statistic (above) and remnant charge imbalance $\langle M_\infty / M_0 \rangle$ (below) are consistent with the absence of MBL in the thermodynamic limit.
In (c), the critical disorder is defined as $\langle r \rangle (W_c) = 0.5$ in the upper panel and $\langle M_\infty / M_0 \rangle (W_c) = 0.2$ in the lower panel.
\label{fig:r_p2_q3_1d}}
\end{figure*}

The $C=1$ subband, on the other hand, behaves qualitatively differently.
The remnant charge imbalance does not increase appreciably with disorder and approaches zero continuously as the system size is increased.
The eigenvalue statistic $\langle r \rangle$ also has no crossing and the value of the critical disorder strength $W_c$ at which it is midway between Poisson and GUE values drifts to infinity with system size.

In Fig.\ \ref{fig:r_p2_q3_1d}, we repeat the same analysis for the case where the projected subbands have Chern numbers $+2$ and $-1$ respectively.
In this case, the lattice commensurability condition constrains the square tori to have a number of unit cells that is a multiple of 3.
This makes fewer sizes amenable to numerical exact diagonalization than in the previous case.


Both subbands tend to delocalize, with the crossover of the $\langle r \rangle$ statistic between GUE and Poisson values drifting rapidly as a function of system size, suggesting absence of MBL in the thermodynamic limit.
The remnant charge imbalance also behaves very similarly, saturating at a very small value, and tending towards zero at all disorder strengths as the system size is increased.
These findings are symptomatic of the absence of MBL in these Chern subbands, as is expected.
Interestingly, there is a small quantitative difference between the $C = -1$ and $C = +2$ subbands in this case: while both show a tendency to delocalize, the $C = +2$ subband delocalizes more easily.
At any given disorder strength $W$ and system size $N_\phi$, the $C=+2$ subband has a larger $\langle r \rangle$ and smaller $\langle M_\infty / M_0 \rangle$ than the $C=-1$ subband.
This is consistent with the theoretical picture in which many-body delocalization is caused by single-particle critical states\cite{Nandkishore2014A}: 
the presence of \emph{two} critical energies\cite{Yang1996,Yang1999,Wan2001} in the $C=+2$ subband makes it quantitatively more robust against localization than the $C=-1$ subband, which has a single critical energy.


\section{Discussion \label{sec:discussion}}

In this paper we have systematically examined the interplay of topology and disorder in the lowest Landau level, with and without interactions.
By providing an explicit recipe to construct potentials that lead to nearly-flat Hofstadter subbands with arbitrary Chern number in the lowest Landau level, we have been able to study various topological subbands individually.
The ability to neglect inter-subband mixing arises from the large bandwidth-to-bandgap ratio ($\approx 10^2$ to $10^4$, depending on the model).

Using exact diagonalization on finite-sized systems,
we analyzed the energy spectrum as a whole, as well as individual many-body eigenstates, in search for signatures of localization.
We find that MBL is absent in topological subbands of the lowest Landau level, in both the one-dimensional and two-dimensional thermodynamic limits. 
Furthermore we find that, at finite size, higher-$|C|$ bands delocalize more strongly than $|C|=1$ bands (such as the whole LLL). 
This corroborates the picture of delocalization being driven by topological ($C \neq 0$) single particle states, which in the presence of interactions act as non-local ``communication channels'' between distant localized single particle orbitals. 
Higher-$|C|$ bands feature more topological states, and are thus even more robust against the localizing effect of disorder.

However, the situation is rather different in topologically trivial LLL subbands.
In principle, such $C=0$ bands have no obstruction to localization since all single-particle states are localized, as we showed in Sec.~\ref{sec:singleparticle}.
While it is unclear if a many-body localized phase is stable in two dimensions in general, our finite-size results suggest that, up to rare region effects which lie beyond the scope of numerical diagonalization, many-body states in $C=0$ subbands of the LLL show several physical signatures of localization -- namely absence of many body level repulsion and persistence of initial conditions.
The results for quasi-1D scaling are fairly unambiguous, and have already been discussed in our previous work\cite{Krishna2019}.
The present work adds weight to the evidence that similar behavior also holds in the two-dimensional thermodynamic limit -- though ultimately further work is needed before a conclusive statement can be made.


\acknowledgments

This work was supported by DOE BES grant DE-SC0002140.

\appendix




\section{Nearly-flat Chern subbands} \label{appendix:flatband}

In this appendix, we discuss details of the engineering of nearly-flat LLL subbands, and prove the validity of the transformation in Eq.~\eqref{eq:Vtrans} for obtaining bands width identical dispersion but tunable Chern number.

We start by deriving the single particle Hamiltonian of an electron in the LLL with a smooth periodic potential, as described in Sec.\ \ref{subsec:cleanpotential}.
This Hamiltonian has discrete translational symmetry, so its spectrum has a Bloch band structure.
A unique feature of projecting the system to the LLL is that the subbands thus formed generically have a topological character described by an integer Chern number.
We also explain how, starting from one set of parameters, a whole family of energetically equivalent yet topologically distinct Hamiltonians may be constructed.

On a rectangular torus of dimensions $L_x \times L_y$ (where $L_x L_y = 2 \pi N_\phi l_B^2$), the LLL wavefunctions form a basis of dimension $N_\phi$.
In the Landau gauge $\mb{A} = - By \hat{\mb{x}}$, the basis wavefunctions are given by
\begin{widetext}
\begin{align}
\psi_n(x, y) = \frac{1}{(\sqrt{\pi} l_B L_x)^{\frac{1}{2}}} \sum\limits_{l=-\infty}^{\infty} e^{i 2 \pi (n + l N_\phi) \frac{x}{L_x}} \exp  - \frac{1}{2 l_B^2} \left[ y - \frac{L_y}{N_\phi}(n + l N_\phi)\right]^2
\end{align}

In this gauge, the magnetic translation operators are \cite{Zak1964}  $t^{(x)}_a = \exp (i \frac{a p_x}{\hbar})$ and $t^{(y)}_b = \exp (i b \frac{p_y - eBx}{\hbar})$.
We wish to study the LLL system with a periodic potential $V(x, y) = V(x+a, y) = V(x, y+b)$, with each unit cell of area $ab$ enclosing $p/q$ flux quanta, with $p$ and $q$ co-prime.
The problem of obtaining the spectrum of energy eigenvalues and the the eigen wavefunctions is simplified by transforming the basis wavefunctions above to a Bloch basis of simultaneous eigenstates of $t^{(x)}_a$, $t^{(y)}_b$ and $V$.
The generic non-commutativity of the translation operators forces us to consider a magnetic unit cell that consists of $q$ primitive unit cells of the periodic potential, such that $[t^{(x)}_{qa}, t^{(y)}_b] = 0$.

The Bloch LLL wavefunctions are \begin{align}
\psi_{\beta, \mb{k}} (x, y) = \frac{\left( 2 \frac{p}{q} \frac{b}{a} \right)^{\frac{1}{4}}}{\sqrt{qab}} \sum_{r = -\infty}^{\infty} e^{i k_y b r} e^{i x (k_x + \frac{2 \pi}{qa}(\beta + rp))} \exp [-\frac{\pi p}{qab} \left( y - \frac{b}{p} \left( \frac{k_x q a}{2 \pi} + \beta + rp \right) \right) ^2],
\end{align}
where the band index $\beta \in \{0, 1, \cdots, p-1 \}$.
The quasimomentum $\mathbf{k}$ is defined through the eigenvalues of the translation operators $t^{(x)}_{qa} \ket{\psi_{\beta, \mb{k}}} = e^{i k_x q a} \ket{\psi_{\beta, \mb{k}}}$ and $t^{(y)}_{b} \ket{\psi_{\beta, \mb{k}}} = e^{i k_y b} \ket{\psi_{\beta, \mb{k}}}$.

In the presence of a periodic potential $V(x,y) = \sum\limits_{m_x, m_y} v_{m_x, m_y} e^{i 2 \pi (m_x x/a + m_y y/b)}$, the degeneracy of the LLL is broken, and we obtain a block-diagonal Hamiltonian at each $\mb{k}$.
Let $H_{\beta, \beta'}(\mb{k}) \equiv \mel{\psi_{\beta, \mb{k}}}{V_\text{1-body}}{\psi_{\beta', \mb{k}}}$, then
\begin{align}
H_{\beta, \beta'}(\mb{k}) &= \sum\limits_{m_x, m_y} v_{m_x, m_y}  e^{-\frac{\pi q}{2 p} (m_x^2 \frac{b}{a} + m_y^2 \frac{a}{b})} \sum\limits_r (-1)^{r m_y} e^{i k_y b r} e^{i 2 \pi \frac{m_y}{p} (\frac{k_x q a}{2 \pi} + \frac{\beta + \beta'}{2})} \delta_{\beta - \beta' + rp + q m_x = 0} \nonumber \\
&= \sum\limits_{\substack{m_x \equiv q^{-1} (\beta' - \beta) \text{ mod } p\\m_y}} \tilde{v}_{m_x, m_y}   e^{-i k_y b (\beta - \beta')/ p} e^{i 2 \pi m_y \beta / p} e^{i \frac{q m_y k_x a - q m_x k_y b}{p} } e^{i \pi q m_x m_y / p},
\label{eq:matrixel}
\end{align}
\end{widetext}
where the LLL-projected Fourier coefficients $\tilde{v}_{m_x, m_y}$ are defined as
 \begin{align}
\tilde{v}_{m_x, m_y} = {v}_{m_x, m_y}e^{-\frac{\pi q}{2 p} (m_x^2 \frac{b}{a} + m_y^2 \frac{a}{b})}
\end{align}
and we have introduced the notation $q^{-1} \text{ mod } p$ to denote the multiplicative inverse of $q$ in $\mathbb Z_p$, i.e. the unique $x\in \mathbb Z_p$ such that $xq \equiv 1 \text{ mod } p$ (this is well defined since $p,q$ are co-prime).

The elements of this matrix can be computed easily, and hence the entire single-particle spectrum over the magnetic Brillouin zone $\mb{k} \in [0, \frac{2 \pi}{qa}] \times [0, \frac{2\pi}{b}]$ can be obtained. 
In Sec.\ \ref{sec:clean}, we first consider the specific case of $p=2$ and $q=1$, which gives us a $2 \times 2$ matrix.
The trace of this matrix is \begin{align}
\Tr H(\mb{k}) & = \sum_{\beta=0,1} H_{\beta\beta}(\mb{k}) \nonumber\\
&= \sum\limits_{\substack{\text{even } m_x, \\ m_y}} \tilde{v}_{m_x, m_y}  e^{\frac{i}{2} (m_y k_x a - m_x k_y b)} \sum_{\beta = 0,1} e^{i\pi m_y \beta} \nonumber \\
& = 2 \sum\limits_{\substack{\text{even } m_x, \\ \text{even } m_y}} \tilde{v}_{m_x, m_y}  e^{\frac{i}{2} (m_y k_x a - m_x k_y b)} 
\end{align}
The matrix can thus be made traceless by setting to zero all Fourier coefficients $v_{m_x, m_y}$ with both $m_x$ and $m_y$ even.
This results in a band structure with symmetry $E_1(\mathbf{k}) = -E_2(\mathbf{k})$.

This single particle Hamiltonian has an interesting feature that enables us to construct the same band structure for two different values of flux per unit cell $p/q$ and $p/q'$, such that $q' \equiv q \mod p$.

This is accomplished by demanding that the LLL-projected Fourier coefficients are the same $\tilde{v}_{m_x, m_y} = \tilde{v}'_{m_x, m_y}$.
As a consequence, the Hamiltonians of the two systems are related by a transformation \begin{align}
e^{-i k_y b \frac{(\beta - \beta')(q-1)}{q}}H_{\beta, \beta'}(\mb{k}/{q}) = e^{-i k_y b \frac{(\beta - \beta')(q'-1)}{q'}} H'_{\beta, \beta'}(\mb{k}/{q'}).
\end{align}
This is a unitary transformation as it may be rewritten in the form \begin{align}
H'(\mb{k}/{q'}) = U H(\mb{k}/{q}) U^\dagger,
\end{align}
where $U$ is a diagonal matrix of complex phases $U_{ll} = e^{i k_y b l \left( \frac{q-1}{q} +\frac{q'-1}{q'} \right)}$.

Since the two Hamiltonians are related by this unitary transformation, the band structures $\{E_1(\mb k), \dots E_p(\mb k)\}$ are identical, up to a rescaling of the magnetic Brillouin zone.

However, the Chern numbers of the subbands are different in the two cases as they are calculated from the eigenstates $\ket{\phi_m(\mb{k})}$ as
\begin{align} 
C_m = \frac{i}{2 \pi} \int \mathrm{d}^2 \mb{k} \ \hat{\mb{z}} \cdot \mel{\nabla_{\mb{k}} \phi_m(\mb{k})}{ \times }{\nabla_{\mb{k}} \phi_m(\mb{k})}.
\end{align}

This transformation is used in the main text to obtain a set of $C = +2, -1$ subbands in a model with $2/3$ magnetic flux quanta per unit cell.

\section{Details on filling interpolation} \label{appendix:nu_interp}

\begin{figure*}[ht!]
\centering
\begin{tabular}{cccc}
\includegraphics[width=0.25\textwidth]{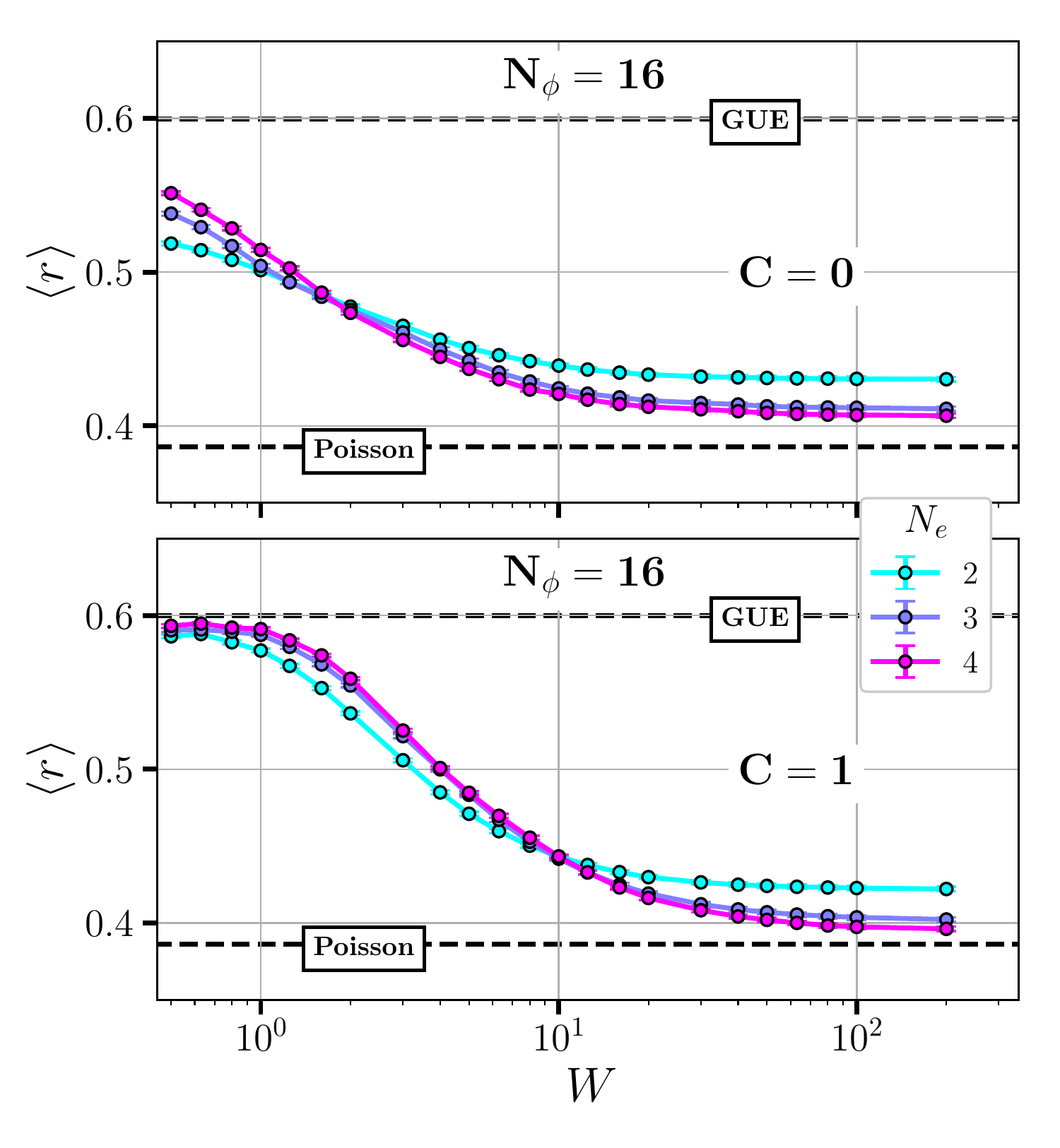} &   \includegraphics[width=0.25\textwidth]{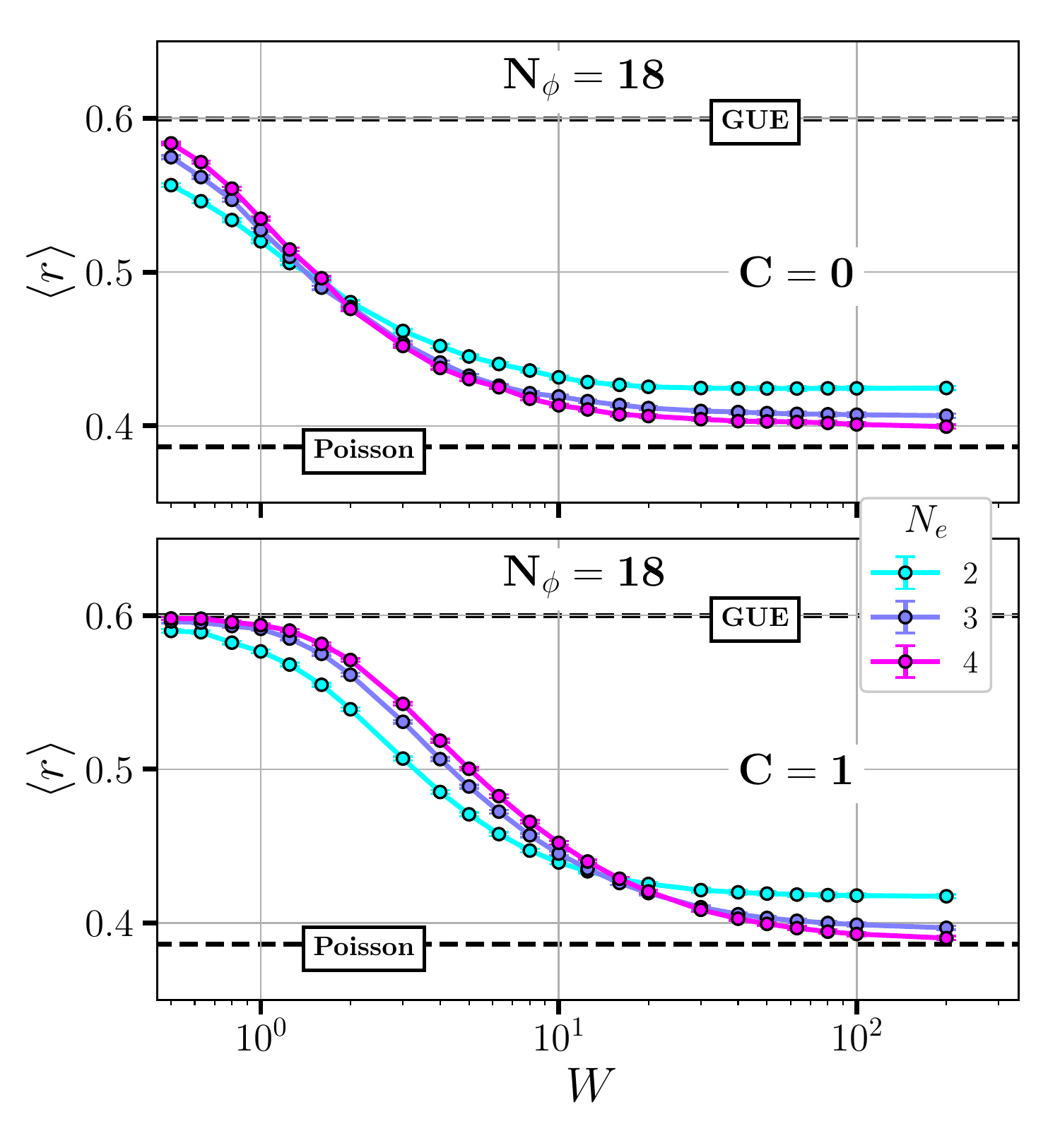} &
\includegraphics[width=0.25\textwidth]{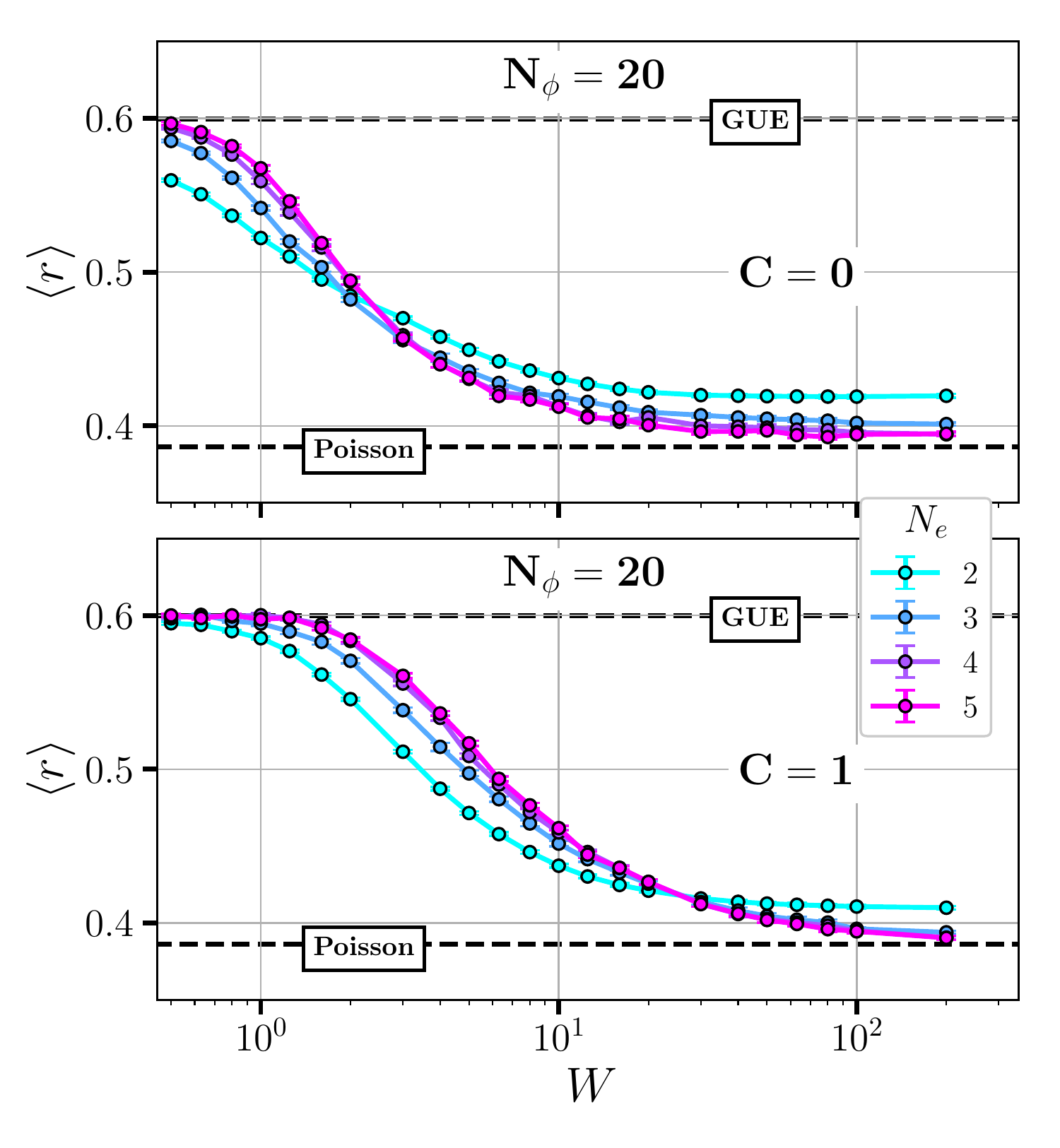} &   \includegraphics[width=0.25\textwidth]{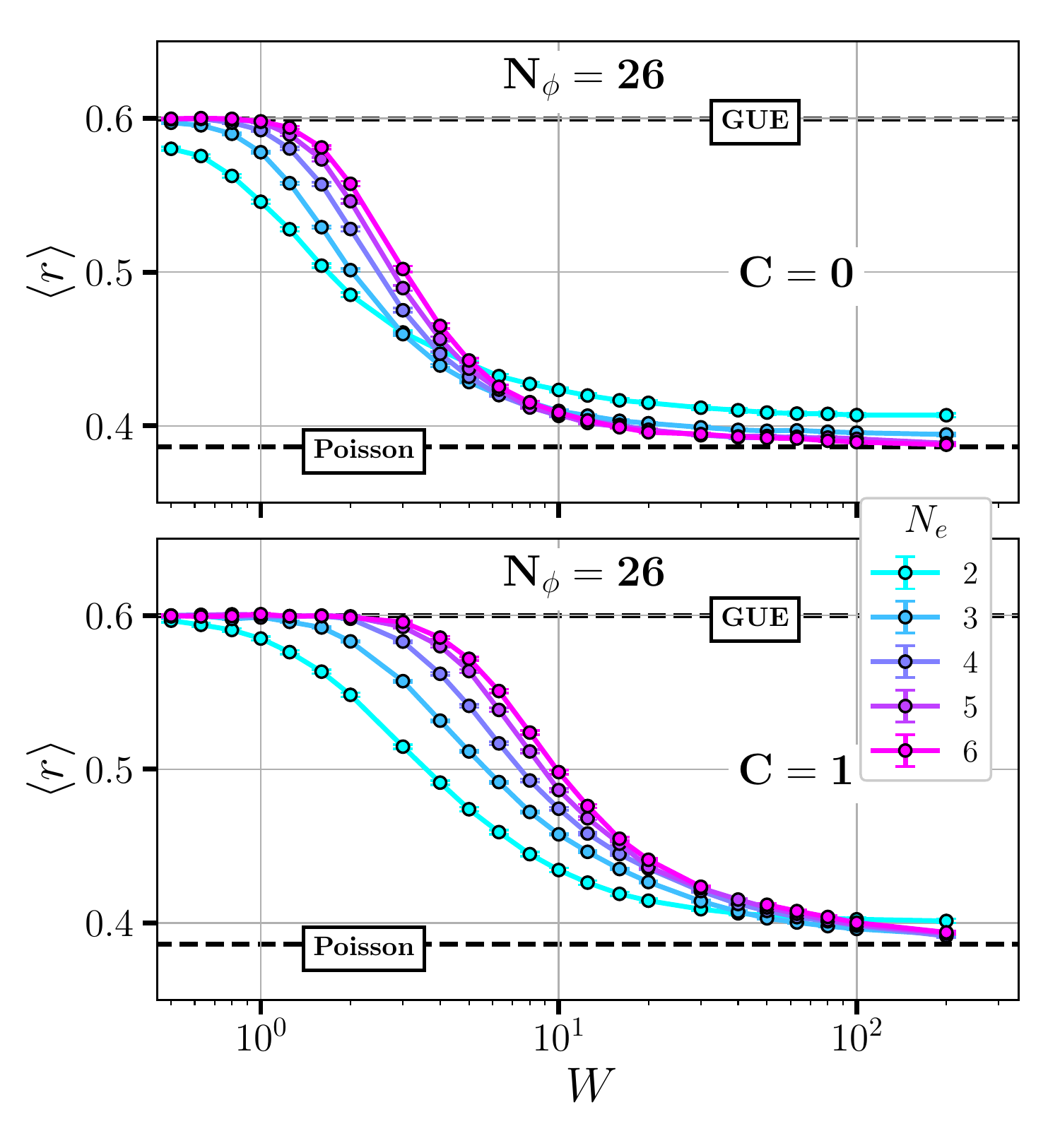} \\ 
\includegraphics[width=0.25\textwidth]{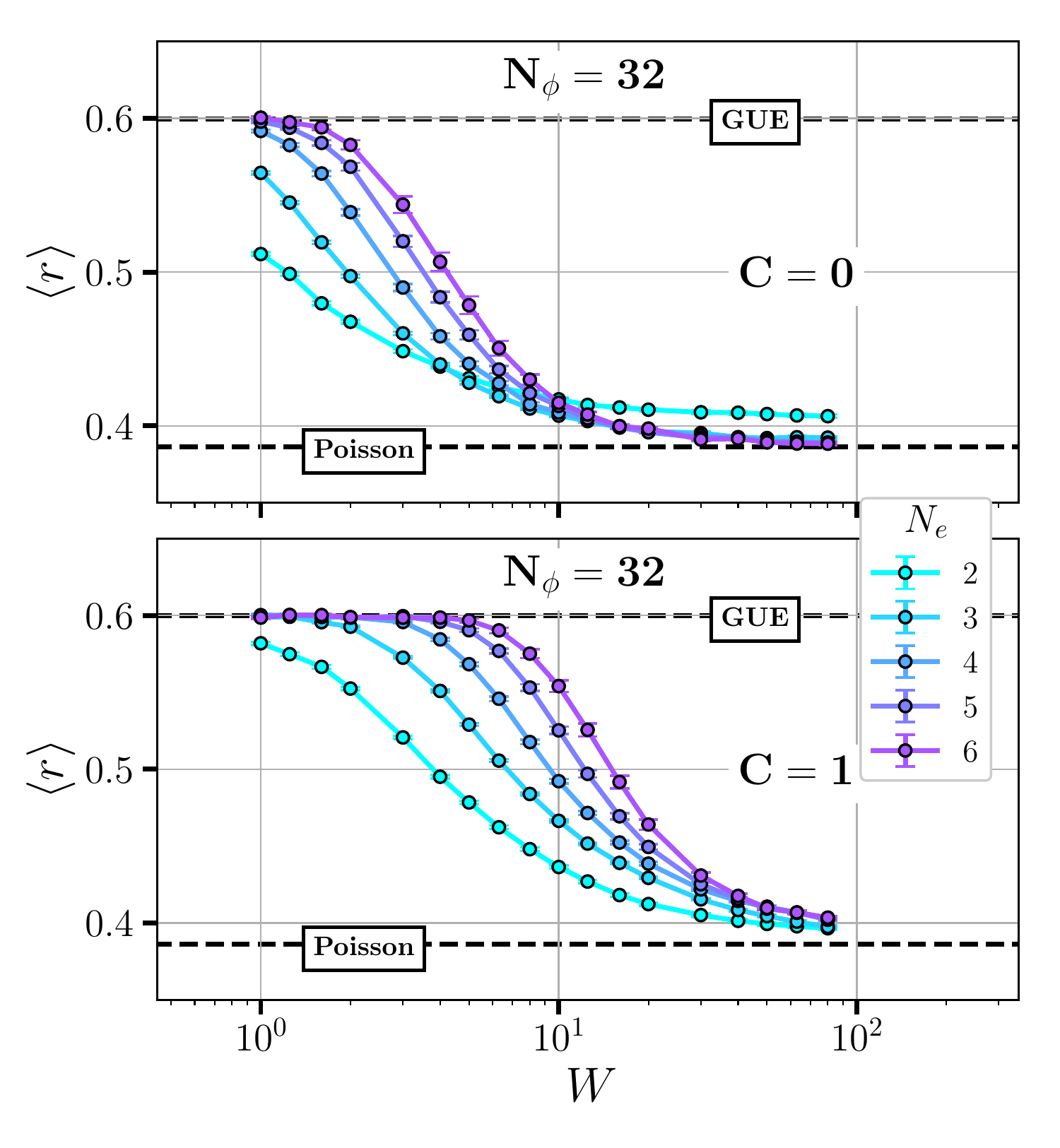} &   \includegraphics[width=0.25\textwidth]{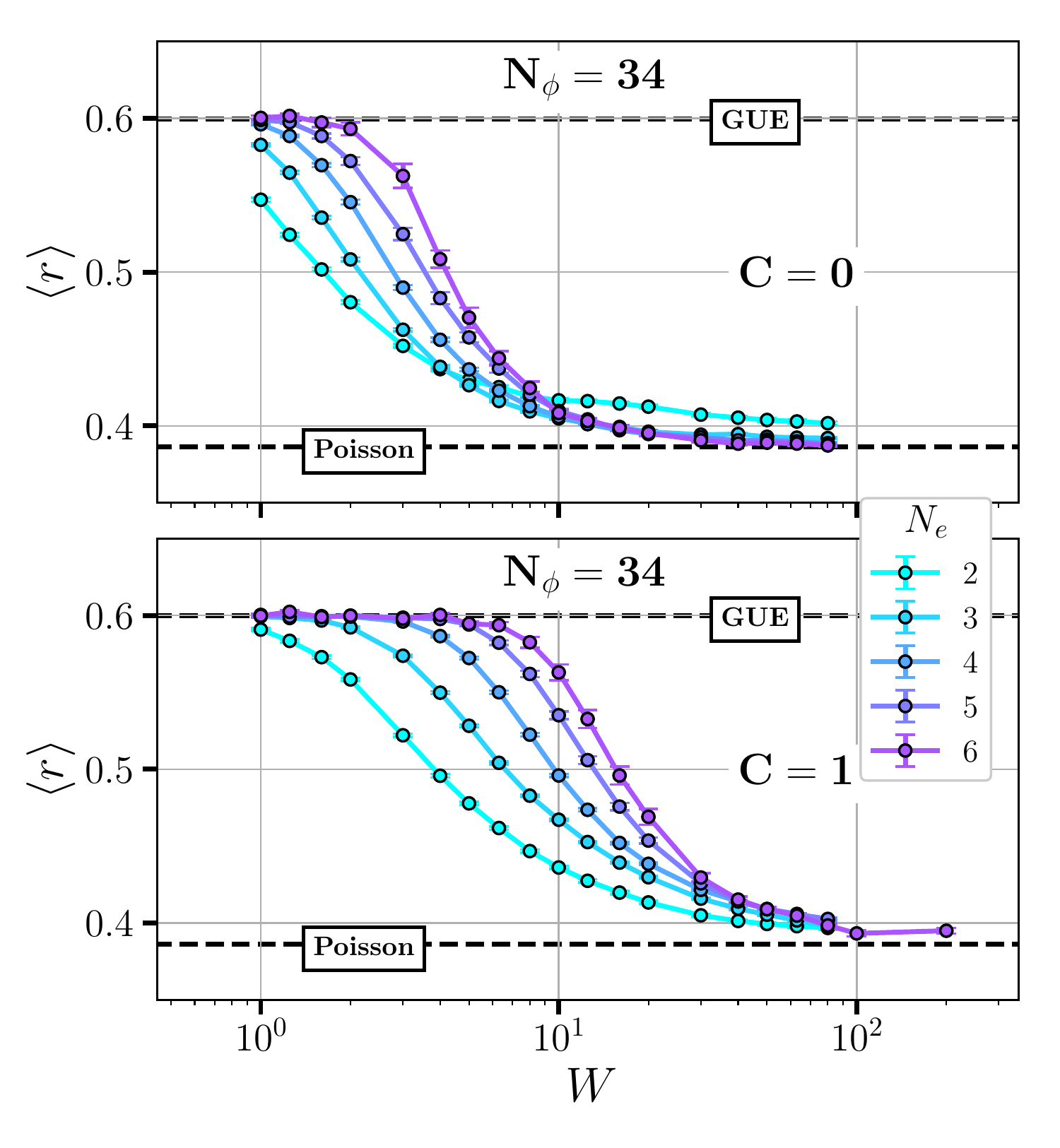} &
\includegraphics[width=0.25\textwidth]{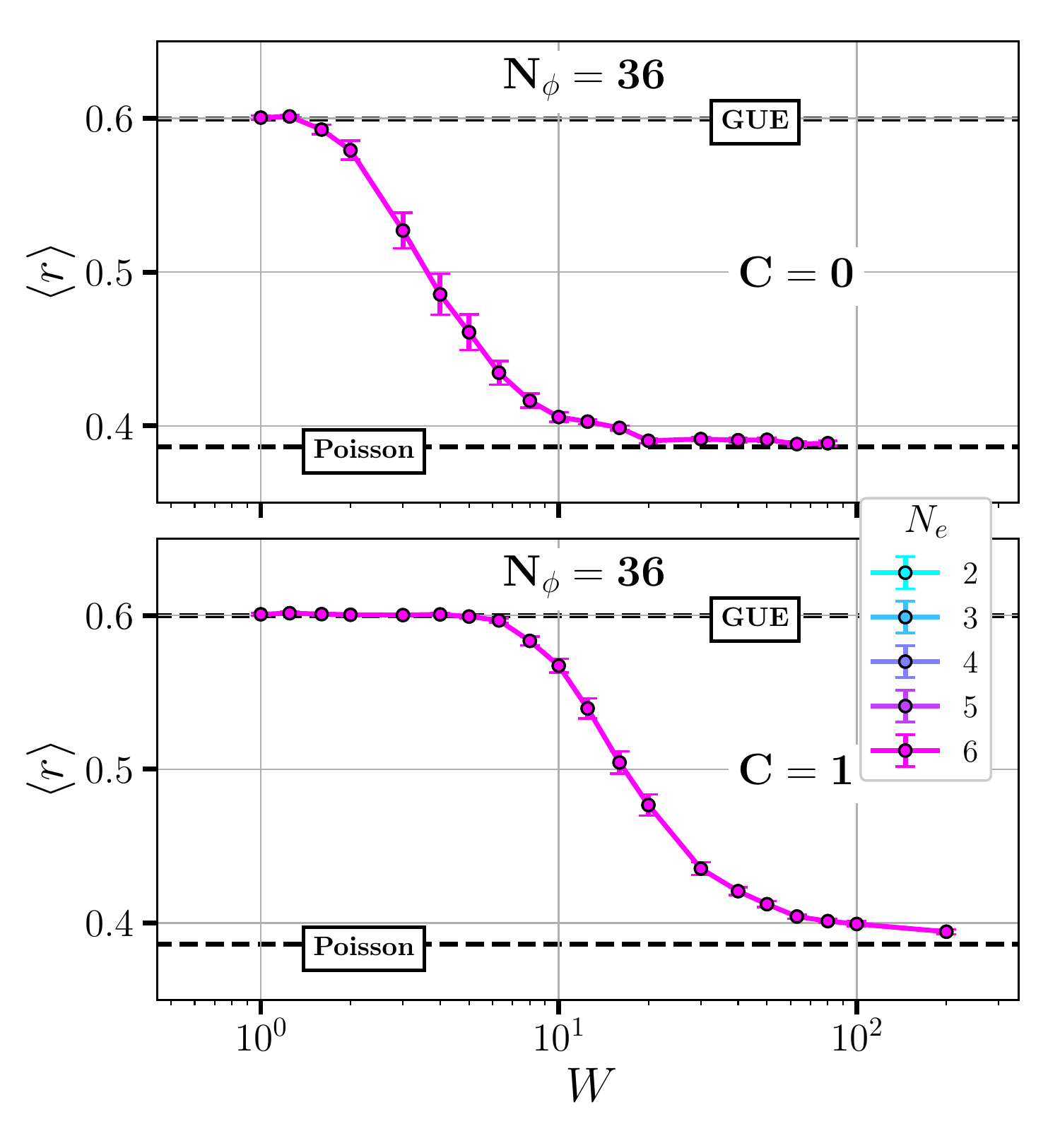} &   
\end{tabular}
\caption{\label{fig:raw_data_r_p2_q1} We plot the raw eigenvalue statistic $\langle r \rangle$ for the eight different dimensions of square tori described in Table \ref{tab:runs_p2_q1}. At each size $N_\phi$, we are able to obtain several different fillings. This enables us to interpolate the data to obtain results at a fixed filling $\nu = 1/3$.}
\end{figure*}

In this appendix, we present the raw data of spectral statistics and charge density imbalance based on which the findings in Sec.\ \ref{subsec:results} are obtained. The periodic potential with two flux quanta per unit cell, as described in Sec. \ref{subsec:cleanpotential}, constrains our system sizes to be $N_\phi = 2(n_1^2+n_2^2)$ with integer $n_1, n_2$ in order to obtain square tori.
For a fixed system size $N_\phi$, we obtain raw data for $\langle r \rangle (W)$ for various fillings $\nu = N_e / 2 N_\phi$ (see Fig.\ \ref{fig:raw_data_r_p2_q1}). 
Since the Hilbert space dimension of the projected problem grows as $\left( \begin{array}{c}
N_\phi/2\\
\nu N_\phi / 2
\end{array} \right)$, limitations on computational power make it impossible to study the exact size scaling at fixed filling $\nu$.

We study the $\langle r \rangle$ statistic as a function of disorder $W$ at different fillings and interpolate the data to obtain an estimate at $\nu = 1/3$.
In Fig.\ \ref{fig:raw_data_r_p2_q1}, we show the raw $\langle r \rangle$ statistic.
At all system sizes, the eigenvalue statistic moves from GUE-like at small disorder to Poisson-like at large disorder.
At a fixed system size $N_\phi$, smaller fillings cross over to the localized Poisson regime more easily.
This is because smaller fillings are more single-particle like and have a smaller Hilbert space dimension compared to fillings near $\nu = 1/2$.
Since the trends are very smooth as a function of system size, we can synthesize a $\langle r \rangle (W)$ curve at any intermediate filling of our choice by linear interpolation between the two nearest fillings at that size.
This enables us to make a comparison of eigenvalue statistics at fixed filling $\nu  = 1/3$ (see Fig.\ \ref{fig:r_p2_q1_2d}).

\begin{figure*}[ht!]
\centering
\begin{tabular}{cccc}
\includegraphics[width=0.25\textwidth]{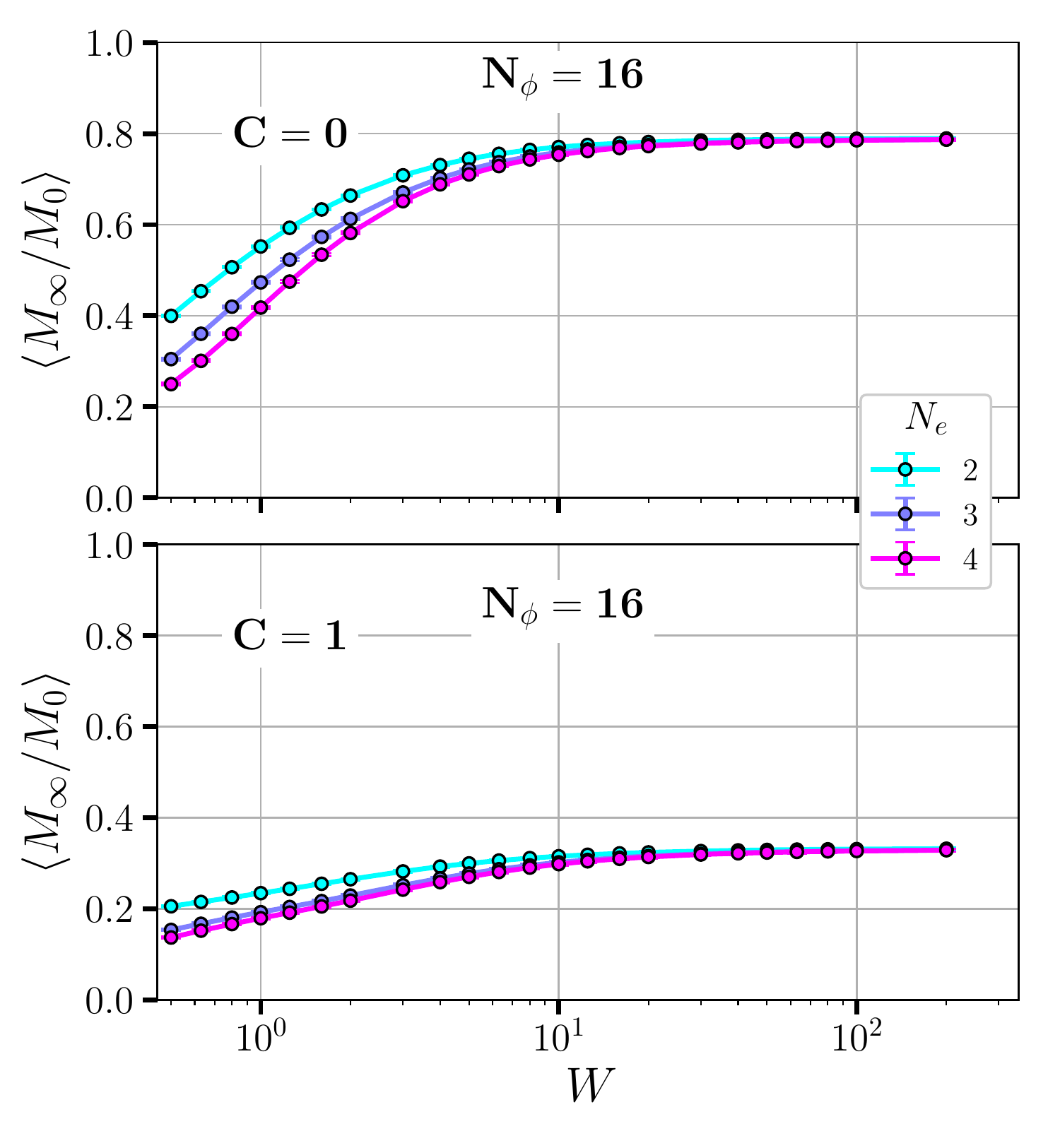} &   \includegraphics[width=0.25\textwidth]{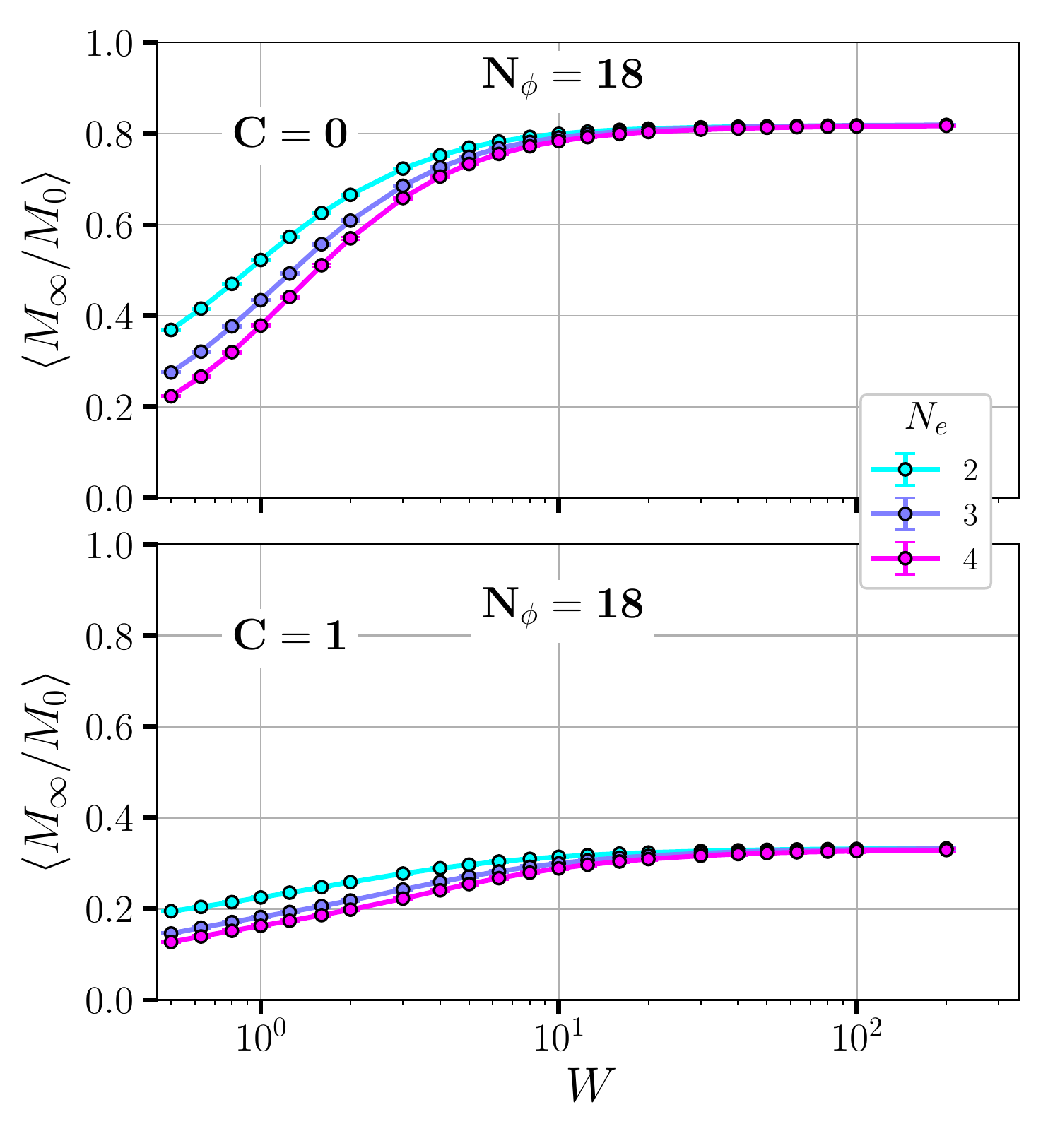} &
\includegraphics[width=0.25\textwidth]{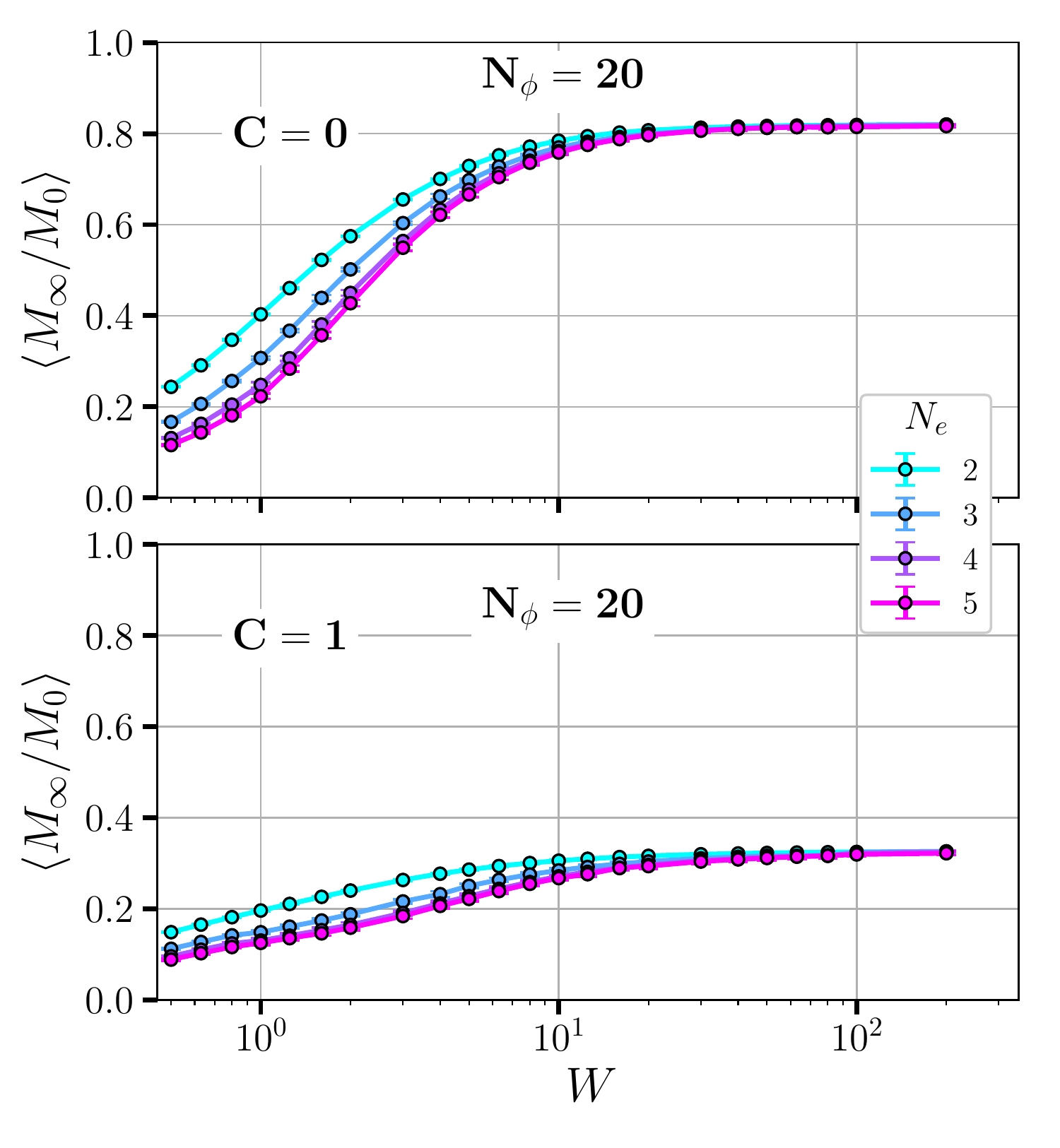} &   \includegraphics[width=0.25\textwidth]{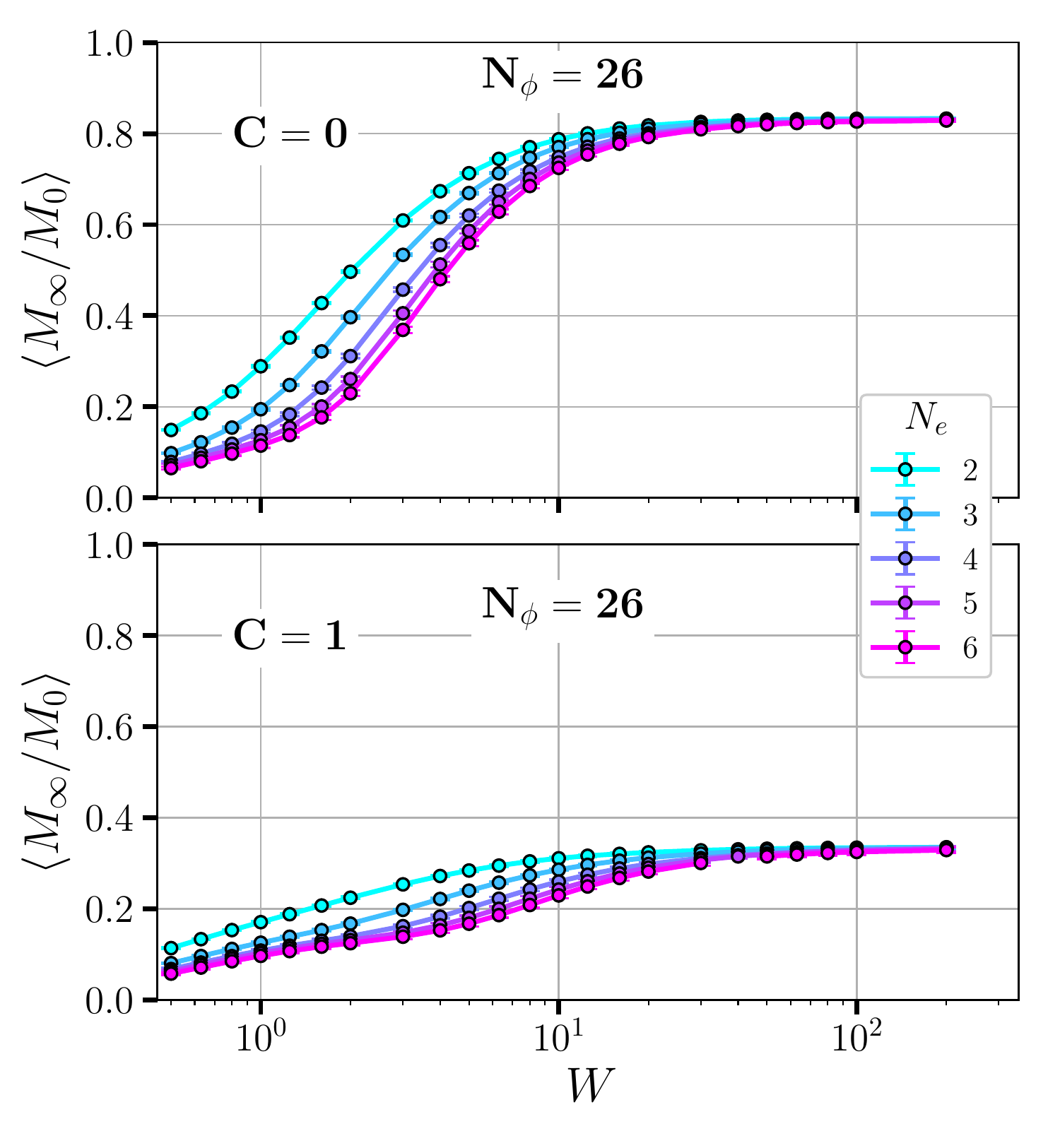} \\ 
\includegraphics[width=0.25\textwidth]{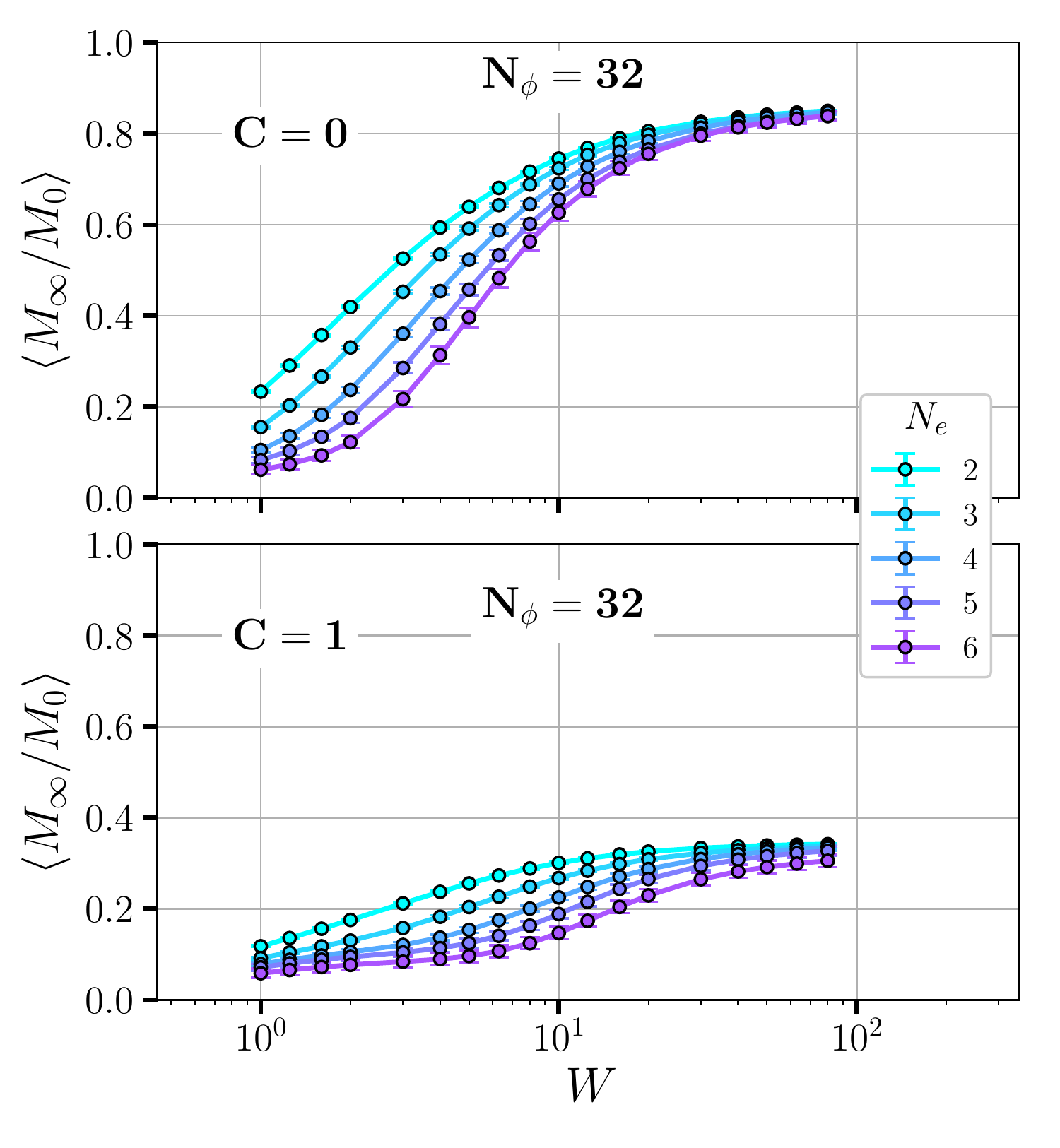} &   \includegraphics[width=0.25\textwidth]{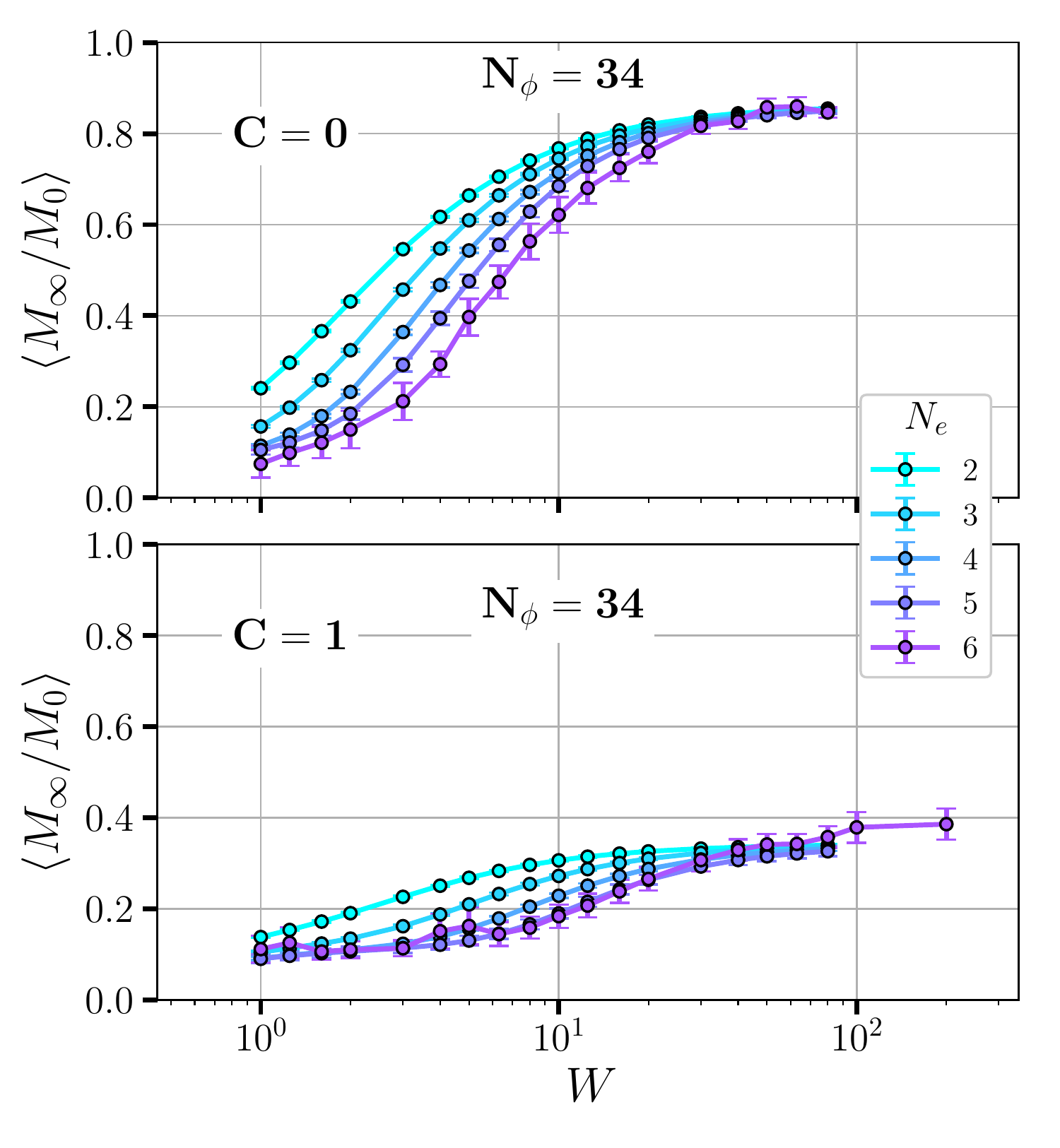} &
\includegraphics[width=0.25\textwidth]{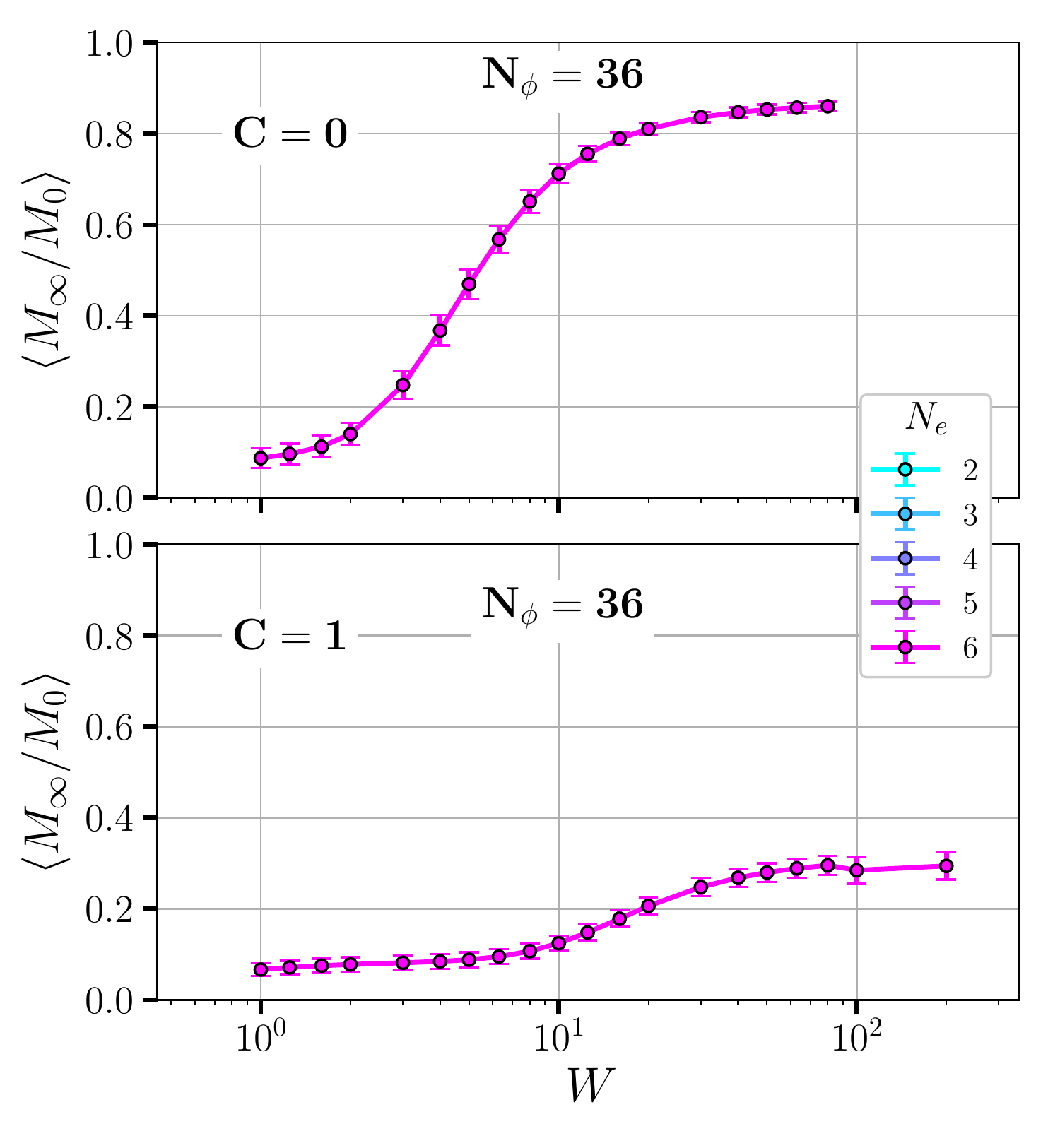} &   
\end{tabular}
\caption{\label{fig:raw_data_M_p2_q1} Similar to Fig.\ \ref{fig:raw_data_M_p2_q1}, we plot the raw charge density imbalance $\langle M_\infty / M_0 \rangle$ for the eight different dimensions of square tori described in Table \ref{tab:runs_p2_q1}.}
\end{figure*}

The same procedure is followed for the charge density imbalance $\langle M_\infty / M_0 \rangle$.
In Fig.\ \ref{fig:raw_data_M_p2_q1}, we show the raw data that is interpolated to obtain the curves in Fig.\ \ref{fig:r_p2_q1_2d}.

\clearpage

\bibliography{qh_loc}

\end{document}